\documentclass[aps,twocolumn,preprintnumbers,amsmath,amssymb]{revtex4}

\makeatletter
\def\@eqnnum{{\normalfont \normalcolor [\theequation]}}
\makeatother

\usepackage{amsmath,amssymb,mathrsfs}
\usepackage{graphicx}
\newcommand{\epsz}{\epsilon_0}
\newcommand{\muz}{\mu_0}

\begin{document}

\title{Magnetic Susceptibility: Further Insights into Macroscopic and Microscopic Fields and the Sphere of Lorentz}
\author{C.~J.~Durrant$^1$, M.~P.~Hertzberg$^1$, 
P.~W.~Kuchel$^{2,}$\footnote{Email: p.kuchel@mmb.usyd.edu.au}}
\affiliation{$^1$School of Mathematics and Statistics, University of Sydney, New South Wales 2006, Australia\\
             $^2$School of Molecular and Microbial Biosciences, University of Sydney, New South Wales 2006, Australia}


\begin{abstract}
To make certain quantitative interpretations of spectra from NMR experiments
carried out on heterogeneous samples, such as cells and tissues, we must be able to
estimate the magnetic and electric fields experienced by the resonant nuclei of atoms in
the sample. Here, we analyze the relationships between these fields and the fields
obtained by solving the Maxwell equations that describe the bulk properties of the
materials present. This analysis separates the contribution to these fields of the molecule
in which the atom in question is bonded, the ``host" fields, from the contribution of all the
other molecules in the system, the ``external" fields. We discuss the circumstances under
which the latter can be found by determining the macroscopic fields in the sample and then
removing the averaged contribution of the host molecule. We demonstrate that the results
produced by the, so-called, ``sphere of Lorentz" construction are of general validity in both
static and time-varying cases. This analytic construct, however, is not ``mystical" and its
justification rests not on any sphericity in the system but on the local uniformity and
isotropy, i.e., spherical symmetry, of the medium when averaged over random microscopic
configurations. This local averaging is precisely that which defines the equations that
describe the macroscopic fields. Hence, the external microscopic fields, {\emph in a suitably averaged
sense}, can be estimated from the macroscopic fields. We then discuss the calculation of the
external fields and that of the resonant nucleus in NMR experiments.
\end{abstract}

\maketitle

\section*{INTRODUCTION}
\subsection*{Overview}
NMR spectroscopy is notable for its contributions to
the study of the chemical and physical properties of
heterogeneous samples including living cells and tissues.
Variations in the magnetic characteristics of a
sample often bring about readily observable changes
in resonance frequency and spectral line shapes, thus
providing unique probes of cellular function (e.g.,
1--4). A knowledge of the physics of systems with
multiple compartments of differing magnetic susceptibility
has already laid the foundation for new sorts of
NMR experiments. The insightful article by Chu et al.
(5) explains some fundamental aspects of contrast
enhancement in magnetic resonance imaging (MRI)
that are brought about by paramagnetic metal--ligand
complexes; and, the comprehensive review in this
journal by Levitt (6) gives an independent explanation
of some key phenomena. Both articles emphasize
the nature and value of the magnetic field ``experienced"
by a nucleus in a magnetically polarizable
medium. They use the theoretical construct of the,
so-called, ``sphere of Lorentz" in their analysis. However,
this theory provoked us into some deeper questions
that seemed to warrant exploration; and, with
this insight came the expectation of a better understanding
of experimental data and new experimental
methods.

The accompanying (preceding) article illustrates
the results from some simple practical NMR experiments
in which samples were chosen in which there
were differences in magnetic susceptibility across the
(micro) boundaries of the heterogeneous samples. The
experiments were conducted on solutions, an emulsion
in the presence of a solution of the same substance,
and a suspensions of red blood cells (RBCs)
made relatively paramagnetic.

\subsection*{Fields in NMR}
NMR spectroscopy is based on the interaction of the
spin and magnetic moment(s) of a nucleus with the
magnetic field in its neighborhood. To simulate NMR
experiments, the magnetic field in the immediate vicinity
of the nucleus in the ``host" atom must be
calculated. The atom itself may be free or bonded as
part of a molecule or in rapid exchange between these
states. In what follows, we refer to the nucleus as
residing in a host molecule with the understanding
that ``molecule" should be interpreted as simply the
host atom when it is not chemically bonded. An NMR
spectrometer has a magnet that is designed to produce
a strong uniform field, which we shall refer to as the
applied field, into which the sample is placed. As a
result of its introduction, the field in and around the
sample is perturbed by the interaction of the field with
the magnetic moments of the molecules in the sample.
Most molecules possess no intrinsic magnetic moment
because the electrostatic binding forces lead to a
net cancellation of the orbital and spin angular momenta
of the electrons. In the presence of an applied
field, the orbital moments precess and generate an
extra component that is aligned opposite to the field;
this is the diamagnetic effect. On the other hand, a few
atoms, ions, and molecules, in which the angular
momenta of the electrons do not cancel, possess an
intrinsic dipole that is much larger than the induced
moment at room temperature. These molecules tend
to adopt the least-energy state in which the intrinsic
dipole moment is parallel to the applied field; this is
the paramagnetic effect. The intrinsic paramagnetic
dipole is much larger than the induced diamagnetic
dipole at room temperatures, so paramagnetic molecules
are often introduced as ``agents" to make a
deliberate modification to the applied field (e.g., 5).
However, the magnetic field experienced by a nucleus
in the host molecule is modified by the fields produced
by all the molecules that are external to the host
in which it is located and by the field produced by the
host molecule itself. The former field is sometimes
called the local field but because it is in the immediate
external environment of the host molecule we shall
refer to it as the external field. The field from the host
molecule we shall refer to as the host field. The
external field is dependent on the composition and
geometry of the whole sample and is a macroscopic
entity, whereas the host field depends solely on the
structure of the particular molecule and is a microscopic
entity. The change in resonance frequency due
to the former is called the bulk magnetic susceptibility
shift (BMS) and that due to the latter is the chemical
shift. (See our preceding article for illustrations of
chemical and BMS shifts that occur in samples in
containers with spherical and cylindrical geometries.)

Both effects are the result of the electromagnetic
properties of charges in motion and are therefore
described by the Maxwell equations (e.g., 7, 8). In
microscopic form, these equations describe exactly
the electric field e and the magnetic induction b produced
by each constituent-charged elementary particle.
Hence, these equations can be used to calculate
the host field and the chemical shift at a nucleus.
However, it is not realistic to solve the array of
Maxwell equations for all the molecules in the sample,
let alone all those in the microscopic system in
the vicinity of the nucleus of interest. Therefore, it is
necessary to invoke a simplification that uses the
Maxwell equations for the macroscopic system, and
then we work down to the microscopic system. The
macroscopic equations describe approximately the
bulk electric field E and the bulk magnetic induction
B, given the constitutive relationships between the
electric field and the electric displacement D, and
between the magnetic induction and the magnetic
field H. These relationships incorporate the effects
due to the paramagnetic and diamagnetic properties of
the bulk medium, described above. But, the macroscopic
field calculated using the macroscopic Maxwell
equations does not provide the average field at a
point where the nucleus resides within the host molecule.
The macroscopic field at any point in fact
contains an average contribution from the host molecule
itself based on the assumption that the point is
distant from the host molecule, not within it. This
average host molecule field contribution must be removed
from the macroscopic field to estimate the
average field across the molecule due to the applied
field and the contributions of all the other molecules,
the external (or local) field. The actual field experienced
by the nucleus is obtained by adding the external
field component and the internal contribution from
the host molecule, namely, the microscopic host field
(see Fig. 1).

\begin{figure}[t]
\includegraphics[width=\columnwidth]{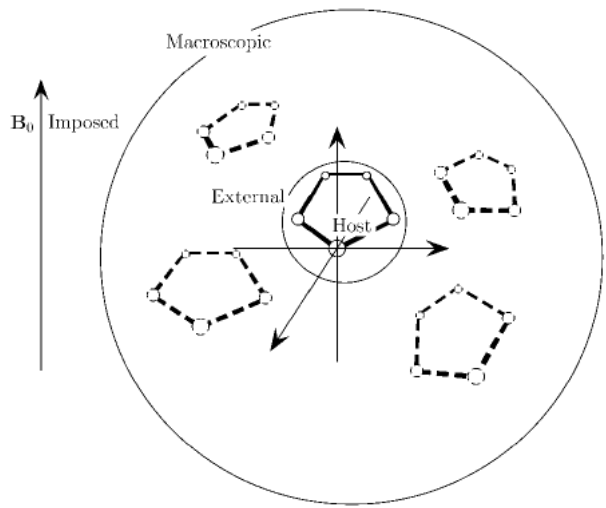}
\caption{Figure 1 Schematic representation of a solution of pentagonal
molecules with nuclei denoted by open discs. The
relative orientation of the molecules is random and the
nucleus of interest is placed at the arbitrarily chosen origin
of a Cartesian coordinate system. The three regions of space
that are relevant to the averaging, or smoothing, process
used in analyzing the fields experienced by the nucleus at
the origin are denoted host, external, and macroscopic.}
\end{figure}

\subsection*{Analytic Strategy}
The commonly used prescription for estimating the
field at the nucleus of interest (the host field) is based
on the idea of the sphere of Lorentz (e.g., 5, 6, 9). It
was introduced in electrostatics by Lorentz (10) but
applies equally well to magnetostatic situations (e.g.,
11). It was first used in the context of NMR by
Dickinson (12). The sphere of Lorentz is a notional
sphere drawn around a nucleus; it is large enough for
all the molecules external to it to be treated as a
macroscopic continuum that is locally uniform.
Within the sphere the host molecule is imagined to
reside in a vacuum surrounded by individual molecules
whose net electromagnetic effect is taken to be
vanishingly small. The effect of introducing a spherical
cavity into a uniform continuous medium is significant
and yet the field arising from it can be readily
calculated; hence, this provides the estimate of the
external field. 

The disadvantage of this approach is that it mixes
the microscopic and macroscopic pictures and the
justification for the field estimates at each level is not
entirely clear. It has thus assumed an almost ``magical"
air; for example, Springer (9) writes ``For an
imaginary object, the sphere of Lorentz produces
amazingly profound real effects.

\subsection*{Aims, Approaches, and Outcomes}
In this article, we set out to demystify the estimation
of the external fields from the macroscopic field(s).
We show that the results obtained by using the sphere
of Lorentz argument are in agreement with our new
more rigorous approach and are of general validity.
This is important because the sphere of Lorentz argument
is simple to apply and can yield a semiquantitative
description of BMS effects in variously shaped
objects.

Our new analysis requires only that the measurement
``polls" a sufficiently large number of molecules
that are in random positions with respect to their
neighbors for the sample average to be equivalent to
the average of a single molecule surrounded by a
randomized medium. Then, local isotropy, i.e., spherical
symmetry, produces the same result as the sphere
of Lorentz construction for a nucleus residing at its
center in the otherwise empty space.

We first summarize the derivation of the expressions
that describe the macroscopic fields, which are
the average of the effect of all molecules present in
the sample; this averaging takes place over macroscopic
length and/or time scales. On the microscopic
scale, all charged particles are taken to be in motion
so that the electric and magnetic properties are coupled,
so we must analyze the full set of Maxwell
equations. It is only on the macroscopic scale that
experiments can realize time scales that are sufficiently
long that the macroscopic equations governing
the magnetic field may be solved independently of
those governing the electric field. At this stage we
confine our discussion to the magnetic field in the
slowly time-varying situation. Next, we examine how
the average fields produced by all the other molecules
at the site of a particular molecule can be estimated by
using the macroscopic equations. The method of solution
of the macroscopic equations is summarized in
the appendix; hence, the calculation of the BMS shift
in NMR experiments is completed.

The detailed treatment of the chemical shift produced
by the internal field of the host molecule at any
of its nuclei is beyond the intended scope of this
article (e.g., 9, 12, 13) but, for completeness, we
provide an estimate of the host field in the spirit of the
treatment of the BMS shift so that the relative magnitudes
of the combined effects can be seen.

\section*{MICROSCOPIC AND MACROSCOPIC FIELDS}

\subsection*{Maxwell Equations}
The Maxwell equations are treated in all standard
texts on electromagnetism (e.g., 7, 8), so we present
them with little discussion.

The microscopic electric and magnetic fields ${\bf e}$ and
${\bf b}$ produced by the moving charges of the submolecular
particles are described by the following equations:
\begin{eqnarray}
\epsz\nabla\cdot{\bf e}&=&\rho,\,\,\,\,\,\,\,\,\,\,\,\,
\nabla\times{\bf e}=-\frac{\partial{\bf b}}{\partial t}\nonumber \\
\nabla\cdot{\bf b}&=&0,\,\,\,\,\frac{1}{\muz}\nabla\times{\bf b}={\bf j}+\epsz\frac{\partial{\bf e}}{\partial t}
\end{eqnarray}
where the constants $\epsilon_0$ and $\mu_0$ are the electrical permittivity
and magnetic permeability of free space,
respectively. We must treat the microscopic electric
and magnetic fields together in a coupled system
because the microscopic charge density $\rho$ and current
density ${\bf j}$ vary on short time scales due to the rapid
motion of the point-like charges.

The Maxwell equations are written here in their
differential form using vector differential operators.
The properties of these operators and their application
in electrostatics are described in (14). But, their integral
form is more readily visualized. Two are obtained
by integrating the two equations on the left over a
volume $V$ and transforming the left sides to integrals
over the surface $S$ of $V$, using Gauss's theorem (12):
\begin{eqnarray}
\iint_S({\bf e}\cdot{\bf n})dS=\frac{1}{\epsz}\iiint_V\rho dV,\,\,\,\,\iint_S({\bf b}\cdot{\bf n})dS=0
\end{eqnarray}
These are illustrated in Fig. 2. 
\begin{figure}[t]
\includegraphics[width=\columnwidth]{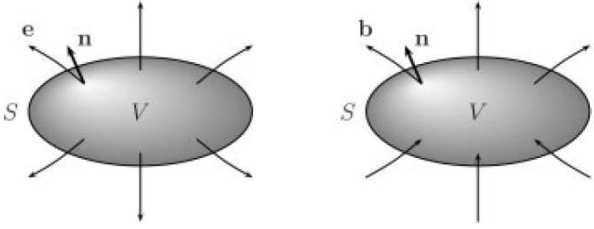}
\caption{Illustrations of Maxwell's laws. Left: The electric
flux is defined as the integral of the component of the
electric field ${\bf e}$ in the direction of the normal ${\bf n}$ to the surface
$S$. It is proportional to the total charge within the volume $V$
enclosed by the surface $S$. Right: The magnetic flux crossing
any surface $S$ enclosing a volume $V$ always vanishes.}
\end{figure}
The unit vector normal
to the surface $S$ at any point is denoted ${\bf n}$ so that the
scalar product, ${\bf b\cdot n}$, is the component of ${\bf b}$ in the
direction normal to the surface. The first equation thus
states that the normal component of ${\bf e}$ integrated over
the surface, the electric flux across the surface, is
proportional to the total charge contained within the
volume $V$. This is Gauss's law. The second equation
states that the total magnetic flux across the surface
surrounding any volume $V$ must vanish. The inward
flux exactly balances the outward flux. This is due to
the empirical fact that there are no sources of magnetism
corresponding to point charges.

The other two forms are obtained by integrating
the two equations on the right in Eq. [1] over a surface
$S$ that is bounded by a closed curve $C$, and then
transforming the left side to line integrals around $C$,
using Stokes's theorem (14):
\begin{eqnarray}
\oint {\bf e}\cdot d{\bf l}&=&-\frac{d}{dt}\iint({\bf b}\cdot{\bf n})dS,\nonumber\\
\frac{1}{\muz}\oint {\bf b}\cdot d{\bf l}&=&\iint({\bf j}\cdot{\bf n})dS
+\epsz\frac{d}{dt}\iint({\bf e}\cdot{\bf n})dS\,\,\,\,\,\,
\end{eqnarray}
Here, $d{\bf l}$ represents a small vectorial increment in
the path around $C$. These are illustrated in Fig. 3. 
\begin{figure}
\includegraphics[width=\columnwidth]{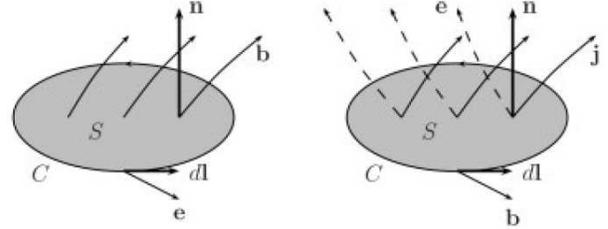}
\caption{Illustrations of Maxwell's laws. Left: The electromotive
force is the integral of the component of the
electric field ${\bf e}$ in the direction of the vector path increment
$d{\bf l}$ around the curve $C$. This integral is equal and opposite to
the rate of change of magnetic flux across the surface $S$.
Right: The corresponding integral of the magnetic field ${\bf b}$
around $C$ is determined by the rate of change of the electric
flux across the surface $S$ and the total current threading the surface.}
\end{figure}
The first equation states that the line integral of the electric
field ${\bf e}$ around a closed path $C$, known as the electromotive
force, is equal and opposite to the rate of
change of the magnetic flux across the surface $S$. This
is Faraday's law of induction. The second equation
states that the line integral of the magnetic field ${\bf b}$
around a closed path $C$ is governed by two quantities.
The first term on the right side is the current crossing
the surface $S$, i.e., threading $C$. This term expresses
Ampere's law. The second term is proportional to the
rate of change of the electric flux across S. This term
was introduced by Maxwell, who called it the displacement
current, and is essential for the existence of
electromagnetic waves.

These equations can be solved by eliminating the
electric field to derive the equation governing the
magnetic field and vice versa. The result in each case
is a wave equation with source terms provided by the
electric charge and current densities:
\begin{eqnarray}
\nabla^2{\bf e}-\frac{1}{c^2}\frac{\partial^2{\bf e}}{\partial t^2}&=&-\frac{1}{\epsz}
\left(-\nabla\rho-\frac{1}{c^2}\frac{\partial{\bf j}}{\partial t}\right)\\
\nabla^2{\bf b}-\frac{1}{c^2}\frac{\partial^2{\bf b}}{\partial t^2}&=&
-\muz\nabla\times{\bf j}
\end{eqnarray}
where $c=1/\sqrt{\epsilon_0\mu_0}$ is the speed of light. The wave
equation can be solved formally in terms of an integral,
over all space and all time, of the sources on the
right sides of these equations. They admit both ``advanced"
and ``retarded" solutions, i.e., one in which
the effect of the source propagates backward in time
with the speed of light and one in which it propagates
forward in time. The advanced solution is discarded to
avoid violating causality. The resulting retarded solution
can then be expressed as an integral over all space
of the contribution from every source at the retarded
time. This time is simply the time at which the electromagnetic
signal, traveling at the speed of light,
must have left the source point to arrive at the selected
point of space ${\bf x}$ where the fields are measured at the
selected time $t$.

This formal integral solution can be written in a
variety of different ways. The Jefimenko form (e.g.,
7) is one of the most illuminating:
\begin{eqnarray}
{\bf e}({\bf x},t)&=&\frac{1}{4\pi\epsz}\iiint\frac{({\bf x}-{\bf x}')\rho'({\bf x}',t')}
{|{\bf x}-{\bf x}'|^3}d^3x'\nonumber\\
&+&\frac{1}{4\pi\epsz}\iiint\bigg{(}\frac{({\bf x}-{\bf x}')(\partial\rho'({\bf x}',t')/\partial t')}
{c|{\bf x}-{\bf x}'|^2} \nonumber\\
&-&\frac{\partial{\bf j}'({\bf x}',t')/\partial t'}{c^2|{\bf x}-{\bf x}'|}\bigg{)}d^3x'\\
{\bf b}({\bf x},t)&=&\frac{\muz}{4\pi}\iiint\frac{{\bf j}({\bf x}',t')\times({\bf x}-{\bf x}')}
{|{\bf x}-{\bf x}'|^3}d^3x'\nonumber\\
&+&\frac{\muz}{4\pi}\iiint\!\frac{(\partial{\bf j}'({\bf x}',t')/\partial t')\times({\bf x}-{\bf x}')}
{c|{\bf x}-{\bf x}'|^2}d^3x'\,\,\,\,\, 
\end{eqnarray}
Without the final term, Eq. [6] is Coulomb's law and,
again without the final term, Eq. [7] is the Biot--Savart
law. In these equations, $t=t-|{\bf x}-{\bf x}'|/c$ is the
retarded time referred to above, and the integration is
taken over all of space. In the steady state, only the
first terms of the right side of Eqs. [6] and [7] appear,
so that the electric field is seen to depend only on the
charge distribution, and the magnetic field depends
only on the current distribution. The size of the second
term relative to the first is 1:($d/c\tau$), where $d$ and $\tau$
 are the characteristic distance and time scales of the
system. If the system changes only slowly (viz., over
time scales long compared with $d/c$, the time for light
to travel a distance $d$), the second terms and the
variation in the retarded time can be neglected; we
call this situation the quasisteady case. At the microscopic
level, the time scale is determined mainly by
the orbital speeds of the electrons that are an order of
magnitude less than c. At the macroscopic level in
NMR experiments, the time scale is governed by
diffusion, i.e., the thermal speeds of the molecules and
these are many orders of magnitude smaller than c.
The quasisteady approximation is, therefore, good in
the latter case, but more care needs to be taken at the
microscopic level.

We now focus on the conceptual steps involved in
proceeding from the microscopic picture in which the
molecular constituents of the material are described as
charged particles in motion in a vacuum, for which
the solutions can be written down exactly, to the
macroscopic picture, which involves systematic averaging
over the sample. Only by doing this is the
relationship between the two pictures revealed. Because
there are several steps that are mathematically
independent but logically connected, we use a notation
that, although somewhat cumbersome, does help
keep track of the steps. 

\subsection*{Temporal and Spatial Averages}
The development presented by Jackson (7) is based
on Russakoff (15) and we follow his approach here:
The Maxwell equations are linear so they can be
averaged over space or time to describe the average
microscopic electric and magnetic fields $\tilde{\bf e}$ and $\tilde{\bf b}$ .
These are expressed in terms of the sources of the
fields, which are the charge density $\tilde{\rho}$ and current
density $\tilde{\bf j}$ that are averaged in the same manner,
namely,
\begin{eqnarray}
\epsz\nabla\cdot\tilde{{\bf e}}&=&\tilde{\rho},\,\,\,\,\,\,\,\,\,\,\,\,\,\,\,\,\,
\nabla\times\tilde{{\bf e}}+\frac{\partial\tilde{{\bf b}}}{\partial t}=0\nonumber \\
\nabla\cdot\tilde{{\bf b}}&=&0,\,\,\,\,\frac{1}{\muz}\nabla\times\tilde{{\bf b}}
-\epsz\frac{\partial\tilde{{\bf e}}}{\partial t}=\tilde{\bf j}
\end{eqnarray}
The form of these equations is precisely that of Eq. [1]
so that the integral forms and the general solution
applied to the averaged fields in the sources are replaced
by the averaged sources.

\subsection*{Significance of Linearity}
The property of linearity enables the fields to be
obtained by averaging the sample point over a volume
in space or interval in time; this is precisely the value
of the field that would be obtained if it were measured
at a fixed point but arose from sources that were
averaged over an equal volume or equal time interval.
As a further consequence of linearity, the average
fields and sources of any system are equal to the sums
of the average fields $\tilde{\bf e}_i$ and $\tilde{\bf b}_i$  due to the average
sources $\tilde{\rho}$ and $\tilde{\bf j}$ representing each of the $N$ molecules
present in the system. Specifically, $\tilde{\bf e}=\sum_{i=1}^N\tilde{\bf e}_i$. Thus,
each average molecular field is described by the Maxwell
equations, Eq. [8], with the corresponding average
molecular sources.

Suppose that the $i$th molecule consists of $n_i$
charged particles, namely, electrons and protons; it
also contains neutrons that generate no electromagnetic
effects per se. We shall not be interested in
following the submolecular motion, so we first average
over a time scale that is long compared with that
of the orbital motion of the electrons but short compared
with that of the motion of some reference point
(usually taken to be the center of mass) of the molecule.
We denote any time-averaged value of a property
$X_i$ with an overbar:
\begin{equation}
\bar{X}_i({\bf x},t)=\int X_i({\bf x},t-\tau)w(\tau)d\tau\nonumber
\end{equation}
where the temporal smoothing function $w(\tau)$ is normalized
so that $\int w(\tau)d\tau =1$ and it vanishes for time
scales longer than that of the submolecular motion.
By treating each submolecular particle as a discrete
point charge, the charge and current density due to the
$j$th charge in the molecule can be expressed in terms
of a delta function (which has, it should be noted,
units of volume$^{-1}$) as
\begin{eqnarray}
\rho_{ij}({\bf x})&=&q_{ij}\delta({\bf x}-{\bf x}_i(t)-{\bf \xi}_{ij}(t)\nonumber\\
{\bf j}_{ij}({\bf x})&=&q_{ij}(\dot{\bf x}_i(t)+\dot{\bf\xi}_{ij}(t))
\delta({\bf x}-{\bf x}_i(t)-{\bf\xi}_{ij}(t))
\end{eqnarray}
Here, ${\bf x}_i(t)$ and $\dot{\bf x}_i(t)$ are the position and velocity of
the center of mass of the $i$th molecule; $q_{ij}$ is the
charge of the $j$th submolecular particle in this $i$th
molecule; and ${\bf \xi}_{ij}(t)$ and  $\dot{\bf \xi}_{ij}(t)$ are the position and
velocity of the $j$th submolecular particle with respect
to its center of mass (see Fig. 4).
\begin{figure}
\includegraphics[width=\columnwidth]{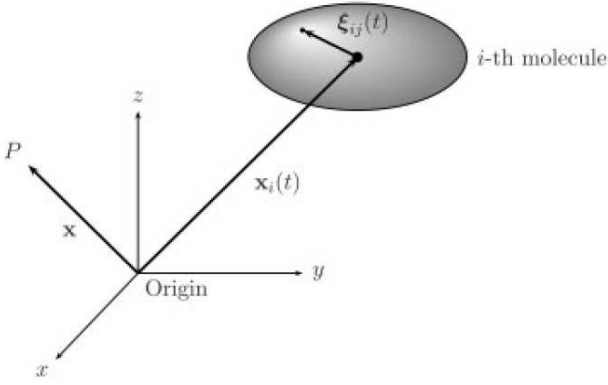}
\caption{Vectors used to specify a general position in
3-space, ${\bf x}$, the position of the center of the $i$th molecule,
${\bf x}_i(t)$, and the position of the $j$th electron in the $i$th atom,
${\bf \xi}_{ij}(t)$, relative to the molecule's center.}
\end{figure}

The delta function $\delta({\bf x}- {\bf x}')$ is a convenient mathematical
device for representing a point particle. This
function vanishes everywhere except at the point ${\bf x}={\bf x}'$, 
yet its integral over any volume containing the
point ${\bf x}'$ is unity. Hence, the quantity $\rho=q\delta({\bf x}-{\bf x}')$
is a charge density such that the total charge $\int\int\int_V\rho dV=0$
if the point charge at ${\bf x}'$ is outside the
volume $V$ and the total charge $\int\int\int_V\rho dV=q$ if the
point charge at ${\bf x}'$ lies within $V$. Likewise, the quantity
${\bf j}= q{\bf v}\delta({\bf x}-{\bf x}')$ is a current density such that the
current $\int\int\int_V{\bf j}dV=0$ if the moving charge at ${\bf x}'$ lies
outside $V$ and the current $\int\int\int_V {\bf j}dV=q{\bf v}$ if the
moving charge at ${\bf x}'$ lies within $V$. Here, ${\bf v}$ is just the
rate at which the point charge is moving; because its
position vector is a function of time, ${\bf x}'= {\bf x}'(t)$, its
velocity is ${\bf v}= d{\bf x}'(t)/dt=\dot{\bf x}'(t)$.

\subsection*{Multipole Expansion}
At a point external to the molecule the inequality $|{\bf x}-{\bf x}_i|>|{\bf\xi}_{ij}|$ 
holds, so we can expand the expressions in
Eq. [9] as series in terms of increasing powers of the
small quantity $|{\bf\xi}_{ij}|/|{\bf x}-{\bf x}_i|$. This procedure leads to
a multipole expansion. Normally, higher-order terms
are small enough to be negligible compared with the
first one or two terms. Truncation of the series after
the second term is referred to as the dipole approximation
and it yields the following expressions for the
total charge and current densities for the $i$th molecule:
\begin{eqnarray}
\bar{\rho}_i({\bf x},t)&\!=&\!q_i\delta({\bf x}-{\bf x}_i(t))-\nabla\cdot(\bar{\bf p}_i(t)
\delta({\bf x}-{\bf x}_i(t))\,\,\,\,\,\,\,\\
\bar{\bf j}_i({\bf x},t)&\!=&\!\frac{\partial}{\partial t}(\bar{\bf p}_i(t)\delta({\bf x}-{\bf x}_i(t))\nonumber\\
&+&\!\nabla\!\times\!(\bar{\bf m}_i(t)\delta({\b x}-{\bf x}_i(t)))\nonumber\\
&+&\! q_i\dot{\bf x}_i(t)
\delta({\bf x}-{\bf x}_i(t))\nonumber\\
&+&\!\nabla\!\times\!(\bar{\bf p}_i(t)\times\dot{\bf x}_i(t)\delta({\bf x}-{\bf x}_i(t)))
\end{eqnarray}
where the total molecular charge $q_i$ (independent of
time), electric dipole moment ${\bf p}_i$, and magnetic dipole
moment ${\bf m}_i$ are defined by
\begin{eqnarray}
q_i &=&\sum_{j=1}^{n_i}q_{ij},\,\,\,\,{\bf p}_i(t)=\sum_{j=1}^{n_i}q_{ij}{\bf\xi}_{ij}(t)\nonumber\\
{\bf m}_i &=&\frac{1}{2}\sum_{j=1}^{n_i}q_{ij}({\bf\xi}_{ij}(t)\times\dot{\bf\xi}_{ij}(t))
\end{eqnarray}
The significance of each of the remaining terms
will be explained below but, in the meantime, these
expressions suffice to convey the idea that when averaged
over time the whole molecule appears at distant
points as if it were a point object. However, it is
conceptually convenient in what follows (although
not mathematically imperative) to restore the finite
extent of the molecule when we deal with its immediate
neighborhood. Thus, we smooth the ith molecule
over a volume $V_i$ that is comparable to the
volume that it occupies. The spatially smoothed quantities
are denoted by a double overbar:
\begin{equation*}
\bar{\bar{X}}_i({\bf x},t)=\iiint\bar{X}_i({\bf x}-{\bf\zeta}',t)h_i({\bf\zeta}')d^3\zeta'
\end{equation*}
where $h_i$ describes the form of the smoothing function
around ${\bf x}$. It vanishes outside $V_i$ and is normalized so
that its integral over $V_i$ is unity.

The Maxwell equations written with these averaged
sources describe fields that vary smoothly over
length and time scales greater than those of the submolecular
structure. However, the sources and fields
still have features on the scale of intermolecular distances.
The macroscopic equations are based on eliminating
this fine structure by being smoothed over a
volume V of space that is sufficiently large to contain
a large number of molecules. These newly smoothed
quantities are denoted by angular brackets:
\begin{equation*}
\langle\bar{\bar{X}}_i({\bf x},t)\rangle=
\iiint\bar{\bar{X}}_i({\bf x}-{\bf\zeta},t)h_V({\bf\zeta})d^3\zeta
\end{equation*}
where $h_V$ is another normalized function, like a
Heaviside step function, that vanishes outside $V$.

The macroscopic averaging process warrants some
discussion. It is implicit in this concept that the result
will describe a physical system that can be measured
with macroscopic equipment and produce repeatable
results, at least within an acceptable error range. As a
result, the macroscopic average must be made over a
volume large enough to ensure that the movement of
the molecular constituents both within this volume as
well as in and out of this volume results in little
change to the average properties. Further, the average
should reflect the results of measurement by different
instruments that may sample the volume in a similar,
but not identical, manner. So, the macroscopic aver-
age should not weight heavily any localized region
within itself. It is therefore logical, as well as expedient
because it simplifies the mathematical treatment,
to take the smoothing function to be uniform over V
so that $h_V=1/V$. Thus, we restrict the integration to
the volume $V$ around ${\bf x}$, writing
\begin{eqnarray}
\langle\bar{\bar{X}}_i({\bf x},t)\rangle=
\frac{1}{V}\iiint\bar{\bar{X}}_i({\bf x}-{\bf\zeta},t)d^3\zeta
\end{eqnarray}

Combining all smoothings, the average charge and
current densities become
\begin{eqnarray}
\langle\bar{\bar{\rho}}_i({\bf x},t)\rangle&\!=&\!q_i H_V({\bf x}-{\bf x}_i(t))\nonumber\\
&-&\nabla\cdot(\bar{\bf p}_i(t)H_V({\bf x}-{\bf x}_i(t))\\
\langle\bar{\bar{\bf j}}_i({\bf x},t)\rangle&\!=&\!\frac{\partial}{\partial t}(\bar{\bf p}_i(t)H_V({\bf x}-{\bf x}_i(t))\nonumber\\
&+&\!\nabla\!\times\!(\bar{\bf m}_i(t)H_V({\b x}-{\bf x}_i(t)))\nonumber\\&+&q_i\dot{\bf x}_i(t)
H_V({\bf x}-{\bf x}_i(t))\nonumber\\
&+&\!\nabla\!\times\!(\bar{\bf p}_i(t)\times\dot{\bf x}_i(t)H_V({\bf x}-{\bf x}_i(t)))
\end{eqnarray}
where
\begin{equation*}
H_V({\bf x})=\frac{1}{V}\iiint_V h_i({\bf x}-{\bf\zeta})d^3\zeta
\end{equation*}
Thus, the charge sources are contributed by the molecular
charge density and spatial variations of the
electric dipole density; the current sources are contributed
by the molecular charge flux, temporal variations
of the electric dipole density, and spatial variations
of the magnetic dipole density and the electric
dipole flux density.

Finally, the macroscopic charge and current
sources are obtained from Eqs. [14] and [15] by
summing over all molecules. Hence,
\begin{eqnarray}
\langle\bar{\rho}({\bf x},t)\rangle&=&-\nabla\cdot{\bf P}({\bf x},t)\\
\langle\bar{\bf j}({\bf x},t)\rangle&=&\frac{\partial{\bf P}({\bf x},t)}{\partial t}
+\nabla\times{\bf M}({\bf x},t)+\nonumber\\
&\nabla&\!\!\times\!\left(\sum_{i=1}^N\bar{\bf p}_i(t)\times\dot{\bf x}_i(t)H_V({\bf x}-{\bf x}_i(t))\right)\,\,\,\,\,
\end{eqnarray}
where the macroscopic polarization ${\bf P}$ and magnetization
${\bf M}$ are defined to be
\begin{eqnarray}
{\bf P}({\bf x},t)&=&\sum_{i=1}^N\bar{\bf p}_i(t)H_V({\bf x}-{\bf x}_i(t)),\nonumber\\
{\bf M}({\bf x},t)&=&\sum_{i=1}^N\bar{\bf m}_i(t)H_V({\bf x}-{\bf x}_i(t))
\end{eqnarray}
At this stage of the analysis, we made the usual
assumptions that the molecular charges $q_i$ and charge
fluxes $q_i \dot{\bf x}_i$ sum to zero over a macroscopic volume, so
the expressions for both the average charge density
and average current density depend solely on the
dipole terms. The second term in the expression for
the average current density is the spatial average of
the (vector) product of the dipole moment and the
center-of-mass velocity of the molecules. If there is
no correlation between the dipole moment and the
velocity (one is microscopic, the other is macroscopic)
then the average of the product is equal to the
product of the averages. In this case, if there is no bulk
motion in the system the average velocity of the
center of mass vanishes and the macroscopic current
density reduces to
\begin{eqnarray}
\langle\bar{\bf j}({\bf x},t)\rangle=\frac{\partial{\bf P}({\bf x},t)}{\partial t}+\nabla\times{\bf M}({\bf x},t)
\end{eqnarray}
This yields the Maxwell equations that describe,
within the approximations detailed above, the macroscopic
fields ${\bf E}$ and ${\bf B}$ in their standard form in the
absence of free charges:
\begin{eqnarray}
\epsz\nabla\cdot{\bf E}&\!=&\!-\nabla\cdot{\bf P},\,\,\,\,
\nabla\times{\bf E}=-\frac{\partial{\bf B}}{\partial t}\nonumber \\
\nabla\cdot{\bf B}&\!=&\!0,\,\,\,\frac{1}{\muz}\nabla\times{\bf B}=\frac{\partial{\bf P}}{\partial t}
+\nabla\times{\bf M}+\epsz\frac{\partial{\bf E}}{\partial t}\,\,\,\,\,\,\,\,\,
\end{eqnarray}
These equations should be contrasted with the
original microscopic form in Eq. [1]. They have exactly
the same structure but the source terms are no
longer discontinuous functions of the microscopic
charges and of the currents produced as they move.
Instead, the sources are continuous functions of the
electric and magnetic dipole densities, which are the
highest-order terms to survive the averaging processes.
However, the general solution of these equations
is the same as for the microscopic equations
given by Eqs. [6] and [7] with the appropriate change
in the expressions for the source terms. In the quasisteady
case these are
\begin{eqnarray}
{\bf E}({\bf x},t)&\!=&\!-\frac{1}{4\pi\epsz}\iiint\frac{({\bf x}-{\bf x}')\nabla\cdot{\bf P}({\bf x}',t)}
{|{\bf x}-{\bf x}'|^3}d^3x'\,\,\,\,\\
{\bf B}({\bf x},t)&\!=&\!\frac{\muz}{4\pi}\iiint\frac{\nabla\times{\bf M}({\bf x}',t')\times({\bf x}-{\bf x}')}
{|{\bf x}-{\bf x}'|^3}d^3x'\nonumber\\
&+&\frac{\muz}{4\pi}\frac{\partial}{\partial t}\iiint\!\frac{{\bf P}({\bf x}',t)\times({\bf x}-{\bf x}')}
{|{\bf x}-{\bf x}'|^3}d^3x'\,\,\,\,\, 
\end{eqnarray}
In practice, these are just formal solutions because
the dipole moment densities are, in general, not prescribed
functions of space and time. Following Maxwell,
we must therefore proceed by introducing the
new fields
\begin{eqnarray}
{\bf D}=\epsz{\bf E}+{\bf P},\,\,\,\,{\bf H}=\frac{\bf B}{\muz}-{\bf M}
\end{eqnarray}
Then, the new Maxwell equations assume the standard
macroscopic forms:
\begin{eqnarray}
\nabla\cdot{\bf B}&\!=&\!0,\,\,\,\,
\frac{\partial{\bf B}}{\partial t}=-\nabla\times{\bf E}\nonumber \\
\nabla\cdot{\bf D}&\!=&\!0,\,\,\,\,\frac{\partial{\bf D}}{\partial t}
=\nabla\times{\bf H}
\end{eqnarray}

We now have four vector field quantities, i.e., 12
scalar quantities, to determine from only eight component
equations. This is impossible without additional
information about how the four fields are related
to one another. But, before considering this
question, we can use the prescription for obtaining the
macroscopic equations to answer the fundamental
question of how such fields can be used to estimate
the fields experienced at the site of a constituent
molecule, i.e., to calculate what we have defined to be
the external fields.

\section*{EXTERNAL FIELDS}

\subsection*{Key Concept}
The key concept in understanding the theoretical construction
derived in this article is: The external fields
at the site of an individual molecule within a sample
differ from the macroscopic fields as calculated above
because the macroscopic field already contains an
averaged contribution to the fields from the molecule
itself. It is obvious that the macroscopically averaged
fields experienced by the kth molecule are found
simply by solving the macroscopic equations, with
sources from the averaged contribution of that molecule
subtracted. It is not obvious, however, that the
actual external fields will always be such fields obtained
by macroscopic spatial averaging. Indeed, in
crystals, they may never be.

\subsection*{Crystals}
In the case of a crystal with molecules arranged in a
regular lattice, there will be no spatial smoothing and
the external fields will be similar for all similar molecules
and could differ greatly from the macroscopic
average. The external fields can only be estimated by
solving the full set of microscopic equations for the
whole lattice; however, this problem is not discussed
further here as it is not pertinent to molecules in
solution.
\subsection*{Amorphous Solids}
In an amorphous solid, with random structure, an
individual molecule also does not experience spatially
smoothed external fields. Consider a group of like
molecules with random arrangements of their neighbors
in a volume $V$. If this volume is used to define the
macroscopic average then the external field experienced
by the group will be similar to the macroscopically
averaged field. The mean field experienced by a
smaller group (in a smaller volume) will, in general,
differ from the macroscopically averaged field because
the system will retain structure on the scale of
intermolecular distances. The evaluation of the external
fields in the latter case poses an intractable problem.

\subsection*{Averaging Process for Fields and Sources}
The question of how to estimate the mean external
field thus hinges on the nature of the averaging process.
Recall that linearity implies that averaging fields
is equivalent to averaging the sources. In a fluid, the
molecular sources move around, so a temporal average
is equivalent to averaging over the locations of the
molecules, i.e., to spatial averaging around a fixed
site. This is the ergodic hypothesis of Boltzmann and
is plausible, although difficult to prove rigorously in
most cases. If a finite volume containing a large
number of molecules with host atoms is sampled,
each such molecule will be surrounded by other molecules
in a series of random realizations. The external
field, averaged over such an ensemble of realizations,
is again equivalent to a spatial average around a fixed
site. The distribution of surrounding sources is then
locally uniform and isotropic within the averaging
volume V. This property characterizes the macroscopic
average and defines how large the necessary
macroscopic averaging volume should be. We shall
consider this case exclusively.

Focus on this averaged spatial distribution of
sources in the neighborhood of an individual molecule,
labeled $k$. To maintain the identity of the molecule,
we assume that it occupies a volume Vk about
the ``center" of the molecule at ${\bf x}_k$; all other molecules
are excluded from this volume. We now smooth the
molecules outside this volume over a spherical volume
$V$ to produce a continuous distribution that maintains
the volume $V_k$ free of sources. As before, we
shall take the smoothing function to be uniform over
$V$. If $V$ does not contain $V_k$, the sources will be
averaged over the whole of $V$ so that $h_V=1/V$ at
points inside $V$ and $h_V=0$ outside. Then, the spatially
averaged sources contributing to Eqs. [14] and
[15] will take the form
\begin{eqnarray}
X({\bf x},t)=\frac{1}{V}\iiint_V\sum_{i=1}^N\!'\bar{\bar{X}}_i({\bf x}-{\bf\zeta},t)d^3\zeta\nonumber\\
X({\bf x},t)=\frac{1}{V}\iiint_V\sum_{i=1}^N\bar{\bar{X}}_i({\bf x}-{\bf\zeta},t)d^3\zeta
\end{eqnarray}
The prime on the sum indicates that the term $i=k$
must be omitted. The second expression follows from
the first because $\bar{\bar{X}}_k$ vanishes throughout $V$. It is precisely
the definition of the macroscopic average $\langle\bar{\bar{X}}\rangle$  in
Eq. [13].

\subsection*{Smoothing Over Sources in the Neighborhood of a Host Molecule}
If we smooth (average) over a volume that includes
any part of $V_k$ we must confine the sources to the
volume $V$ minus $V_k'$, which is that part of $V_k$ within $V$
(see Fig. 5). 
\begin{figure}
\includegraphics[width=\columnwidth]{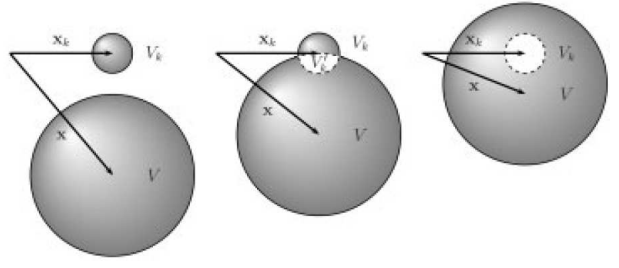}
\caption{Molecular volumes $V_k$, centered at ${\bf x}_k$, and the
smoothing volumes $V$, centered at ${\bf x}$. Left: The molecule lies
wholly outside $V$. Right: The molecule lies wholly inside.
Center: That part of the molecular volume lying inside V is
denoted $V_k'$.}
\end{figure}
This volume depends on the position of
the center, ${\bf x}$, of the smoothing sphere relative to the
center of the molecule ${\bf x}_k$. The normalization now
gives $h_V=1/(V-V_k')$ and the average sources are
\begin{eqnarray*}
X({\bf x},t)=\frac{1}{(V-V_{k}')}\iiint_{V-V_{k}'}\sum_{i=1}^N\!'\bar{\bar{X}}_i({\bf x}-{\bf\zeta},t)d^3\zeta\nonumber\\
X({\bf x},t)=\frac{1}{(V-V_{k}')}\iiint_{V-V_{k}'}\sum_{i=1}^N\bar{\bar{X}}_i({\bf x}-{\bf\zeta},t)d^3\zeta
\end{eqnarray*}
because $\bar{\bar{X}}_k$
vanishes throughout $V-V_k'$. This does
not match the definition of the macroscopic average
but it can be further expressed as
\begin{eqnarray*}
X({\bf x},t)=\frac{1}{(V-V_{k}')}\Bigg{(}
\iiint_{V}\sum_{i=1}^N\bar{\bar{X}}_i({\bf x}-{\bf\zeta},t)d^3\zeta\nonumber\\
-\iiint_{V_{k}'}\bar{\bar{X}}_k({\bf x}-{\bf\zeta},t)d^3\zeta\Bigg{)}
\end{eqnarray*}
Now, the first integral is the total source $Q_V$ within
$V$. Because $V$ is the macroscopic volume, this
source is given by Eq. [25] as $Q_V=V\langle\bar{\bar{x}}\rangle$. The
molecules will be randomly distributed over this
volume with an average number density $1/V_m$,
where $V_m$ is the average molecular ``volume" so we
can define the average source associated with each
molecule to be
\begin{equation*}
Q_m=\frac{V_m}{V}Q_V=V_m\langle\bar{\bar{X}}\rangle
\end{equation*}
unless the total source vanishes, i.e., $Q_V= V\langle\bar{\bar{X}}\rangle=0$.
The second integral is the source due to the $k$th
molecule within $V_k'$. By definition, the source within $V_k$ is
\begin{equation*}
Q_k=\iiint_{V_k}\bar{\bar{X}}_k({\bf x}-{\bf\zeta},t)d^3\zeta
\end{equation*}
If $\bar{\bar{X}}_k$ is uniform within $V_k$, then it follows that
\begin{equation*}
\iiint_{V_k'}\bar{\bar{X}}_k({\bf x}-{\bf\zeta},t)d^3\zeta=\frac{V_k'}{V_k}Q_k
\end{equation*}
If $\bar{\bar{X}}$ is not uniform, we can always write,
\begin{equation*}
\iiint_{V_k'}\bar{\bar{X}}_k({\bf x}-{\bf\zeta},t)d^3\zeta=f_X({\bf x}-{\bf x}_k)\frac{V_k'}{V_k}Q_k
\end{equation*}
where $f_X$ is some smooth function of position with
respect to the center of the molecule and is a function
of the structure or shape of the kth molecule. When all
the molecules are of the same type $Q_k= Q_m$; otherwise,
we write $Q_k=\alpha_k Q_m$, where $\alpha_k$ is a proportionality
constant that depends on the other molecular
structure(s). Then, substitution yields
\begin{equation*}
X({\bf x},t)=\frac{V}{(V-V_k')}\langle\bar{\bar{X}}-\alpha_k f_X\frac{V_m}{(V-V_k')}\frac{V_k'}{V_k}
\langle\bar{\bar{X}}\rangle
\end{equation*}
if $\langle\bar{\bar{X}}\rangle\ne 0$. This case describes the contributions of
the molecular dipoles ${\bf p}_k$ and ${\bf m}_k$ to the sources in
Eqs. [14] and [15]. When $\langle\bar{\bar{X}}\rangle = 0$,
\begin{equation*}
X({\bf x},t)=-f_X\frac{V_k'}{(V-V_k')V_k}Q_k
\end{equation*}
This case describes the contribution of the possible
molecular charge $q_k$ to Eq. [14] because
\begin{equation*}
Q_k=q_k\iiint_{V_k}h_k({\bf x}-{\bf x}_i-{\bf\zeta})d^3\zeta=q_k
\end{equation*}
At locations within $V_k$ we must, of course, set $X({\bf x},t)=0$.

\subsection*{Alternative Representation of the Sources}
The results of the previous section can be put in an
alternative, perhaps more illuminating, form by noting
that they differ from the sources for the macroscopic
fields only within $V_k$ or when $V$ includes some part of
$V_k$. The external fields are therefore obtained from the
macroscopic fields ${\bf E}$ and ${\bf B}$ by removing the fields due
to the contribution of the $k$th molecule to the sources
within $V_k$, and around $V_k$. We call these the self-fields.
They are produced by the sources, as follows:
\begin{widetext}
\begin{eqnarray*}
X({\bf x},t)=\Bigg{\{}
\begin{array}{ll}
\langle\bar{\bar{X}}\rangle & \mbox{if ${\bf x}$ is within $V_k$}\\
\mathscr{S}_X({\bf x}-{\bf x}_k)\langle\bar{\bar{X}}\rangle\,\, & \mbox{if $V$ about ${\bf x}$ contains
any part of $V_k$}\\
0 & \mbox{if $V$ about ${\bf x}$ is wholly outside $V_k$}
\end{array}
\end{eqnarray*}
or, if $\langle\bar{\bar{X}}\rangle=0$,
\begin{eqnarray*}
X({\bf x},t)=\Bigg{\{}
\begin{array}{ll}
0 & \mbox{if ${\bf x}$ is within $V_k$}\\
\mathscr{S}_Q({\bf x}-{\bf x}_k)Q_k\,\,\, & \mbox{if $V$ about ${\bf x}$ contains
any part of $V_k$}\\
0 & \mbox{if $V$ about ${\bf x}$ is wholly outside $V_k$}
\end{array}
\end{eqnarray*}
\end{widetext}
The shape functions,
\begin{eqnarray}
\mathscr{S}_X &=&\frac{V_k'}{(V-V_k')}\left(\frac{\alpha_k f_X V_m}{V_k}-1\right),
\nonumber\\
\mathscr{S}_Q &=&\frac{f_Q V_k'}{(V-V_k')V_k},
\end{eqnarray}
contain all the effects of the particular structure of the
kth molecule. The other quantities are all macroscopic
quantities.

The expressions for the self-fields then result from
solving the Maxwell equations with charge densities:
\begin{widetext}
\begin{eqnarray*}
\rho({\bf x},t)=\Bigg{\{}
\begin{array}{ll}
-\nabla\cdot{\bf P} & \mbox{if ${\bf x}$ is within $V_k$}\\
\mathscr{S}_Q q_k-\nabla\cdot(\mathscr{S}_p{\bf P})\,\, & \mbox{if $V$ about ${\bf x}$ contains
any part of $V_k$}\\
0 & \mbox{if $V$ about ${\bf x}$ is wholly outside $V_k$}
\end{array}
\end{eqnarray*}
and current densities: 
\begin{eqnarray*}
{\bf j}({\bf x},t)=\Bigg{\{}
\begin{array}{ll}
\partial{\bf P}/\partial t+\nabla\times{\bf M} & \mbox{if ${\bf x}$ is within $V_k$}\\
\partial(\mathscr{S}_p{\bf P})/\partial t+\nabla\times(\mathscr{S}_M{\bf M})
+\mathscr{S}_Q q_k\dot{\bf x}+\nabla\times(\mathscr{S}_p{\bf P}\times\dot{\bf x})
\,\, & \mbox{if $V$ about ${\bf x}$ contains
any part of $V_k$}\\
0 & \mbox{if $V$ about ${\bf x}$ is wholly outside $V_k$}
\end{array}
\end{eqnarray*}
\end{widetext}

\subsection*{Prescription for External Fields}
The previous section provides a prescription for evaluating
the external field experienced by an individual
molecule due to a surrounding randomized configuration
of other molecules. The equations may be
solved if the structure and dynamics (position, orientation,
and velocity) of the molecule are known. In
practice, the dynamic state of individual molecules
will not in general be known, so this formalism has
limited utility. However, in NMR experiments the
signal is generated by a large number of nuclei residing
in molecules, all in different dynamic states.
Therefore, we can perform another averaging: This
time it is done over the velocities and orientations of
the host molecule. If there is no net flux of the
molecules in question, i.e., there is negligible bulk
motion, the velocity-dependent terms will average to
zero. If we also average over all orientations then the
newly averaged sources must be distributed with
spherical symmetry. The host molecule is thereby
replaced by one that is spherically symmetrical, as in
Fig. 6. 
\begin{figure}
\includegraphics[width=\columnwidth]{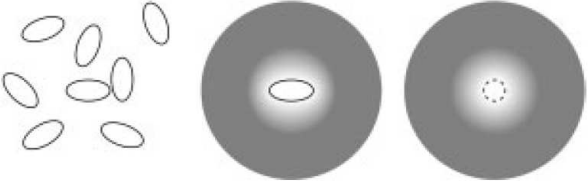}
\caption{Successive smoothing processes. Left: The host
molecule is surrounded by other molecules in random positions
and orientations. Middle: The surrounding molecules
have been smoothed into a continuous source distribution
outside the host molecule. Right: The orientation of the host
molecule has been averaged, resulting in a spherically symmetrical
distribution of sources outside a spherical ``exclusion"
volume.}
\end{figure}
The molecular volume $V_k$ will then be a sphere
and all spatially dependent factors in the shape functions $\mathscr{S}$ 
will depend only on distance $r=|{\bf x}-{\bf x}_k|$
from the center at ${\bf x}_k$.	

If we recall that the point ${\bf x}_k$ is some reference
point in the molecule, not necessarily the center of
mass, it is clear that we can now choose that point to
be the location of the nucleus in the host molecule.
When the orientations of the molecule are averaged
about this point, the averaged sources due to all the
other molecules become spherically symmetrical
about the nucleus. The average fields experienced by
that nucleus are those due to the spherically symmetrical
source distributions, evaluated at the center, i.e.,
at ${\bf x}={\bf x}_k$.

We can now construct the sources explicitly. Letting
$R$ be the radius of $V$, and $R_k$ the radius of $V_k$, the
charge densities are given by
\begin{equation*}
\rho({\bf x},t)=\Bigg{\{}
\begin{array}{ll}
-\nabla\cdot{\bf P} & \mbox{if $r<R_k$}\\
\mathscr{S}_Q(r) q_k-\nabla\cdot\mathscr{S}_p(r){\bf P} & \mbox{if $R_k<r<R+R_k$}\\
0 & \mbox{if $R+R_k<r$}
\end{array}
\end{equation*}
and the current densities are given by
\begin{eqnarray*}
{\bf j}&&\!\!\!\!\!\!\!({\bf x},t)=\nonumber\\
\Bigg{\{}&&\!\!\!\!\!\!\!
\begin{array}{ll}
\partial{\bf P}/\partial t+\nabla\!\times\!{\bf M} & \mbox{if $r<R_k$}\\
\partial(\mathscr{S}_p(r){\bf P})/\partial t+\nabla\!\times\!(\mathscr{S}_M(r){\bf M})
 & \mbox{if $R_k<r<R+R_k$}
\\0 & \mbox{if $R+R_k<r$}
\end{array}
\end{eqnarray*}
where ${\bf P}$ and ${\bf M}$ are the macroscopic polarization and
magnetization around ${\bf x}={\bf x}_k$.

\subsection*{Isolated Molecules}
Consider the idealized case in which there is a single
molecular species (so $\alpha_k=1$ and $q_k=0$) and set
$V_k=V_m$ and $f_X=0$ so that each molecule occupies
the same exclusive spherical volume that is equal to
the molecular volume. Then, the shape functions vanish,
leaving only the uniform sources ${\bf P}$ and ${\bf M}$ within
$V_k$. Substitution into the quasisteady expressions Eqs.
[21] and [22] yields explicit estimates of the selffields
at ${\bf x}={\bf x}_k$ that are given by
\begin{eqnarray}
{\bf E}_{\mbox{\tiny{self}}}=-\frac{1}{3\epsz}{\bf P}({\bf x}_k,t),\,\,\,\,
{\bf B}_{\mbox{\tiny{self}}}=\frac{2\muz}{3}{\bf M}({\bf x}_k,t)
\end{eqnarray}

The contribution from the polarization ${\bf P}$ to ${\bf B}_{\mbox{\tiny{self}}}$
integrates to zero at the center of the sphere. This does
not occur at other points within the sphere. However,
in the strictly time-independent case the fields are
uniform within a sphere with uniform polarization and
magnetization so Eq. [27] gives the estimate of the
static self-fields at all points within the sphere. This
case reproduces the result obtained from the sphere of
Lorentz construction, even though the spheres are
differently conceived and are of different size. The
contradictions inherent in this simplistic view of the
sphere of Lorentz are apparent, however, so we pursue
the rigorous treatment.

\subsection*{Rigorous Treatment}
If we wish to identify values averaged over V in the
general case with the macroscopic averages, we must
take $V\gg V_k$. The macroscopic sources are then
uniform throughout $V_k$ and its surroundings, within
the much larger volume $V$ about the molecule. Thus,
in the surrounding shell only the shape functions vary
and then only with respect to the radial coordinate $r$.
The importance of this property was noted by (6);
substitution of these forms in Eqs. [21] and [22]
produces exactly the same estimates as those in Eq.
[27] for the fields at the center of Vk. Away from the
center, the time-varying polarization will again contribute
to Bself. In the strictly time-independent case,
the fields will again be uniform throughout $V_k$, so
these estimates will apply at all points within $V_k$.

The external fields experienced at any position ${\bf x}$
within an {\emph averaged} molecule in {\emph averaged} surroundings
are the macroscopic fields less the self-fields. In
the static case, these are
\begin{eqnarray}
{\bf E}_{\mbox{\tiny{ext}}}({\bf x})={\bf E}({\bf x})+\frac{1}{3\epsz}{\bf P}({\bf x}),\nonumber\\
{\bf B}_{\mbox{\tiny{ext}}}({\bf x})={\bf B}({\bf x})-\frac{2\muz}{3}{\bf M}({\bf x})
\end{eqnarray}
In the time-varying case, these expressions are exact
at the center of the symmeterized molecule and approximate
the fields close to the center.

The most significant conclusion, however, follows
from the fact that the nucleus of the host molecule is
placed, by construction, at the center of the symmeterized
distributions. The expressions in Eq. [28]
therefore give the average fields experienced by the
nucleus as a result of the surrounding molecules exactly,
even if the fields are not static.

\subsection*{Alternative Expression for the External Field}
An alternative form for the expression for external
fields can be obtained by recognizing that our assumptions
allow us to rewrite Eq. [18] as
\begin{eqnarray}
{\bf P}=N\bar{\bf p}_m,\,\,\,\,{\bf M}=N\bar{m}_m
\end{eqnarray}
where $N$ is the number density of molecules and $\bar{\bf p}_m$
and $\bar{\bf m}$ are the mean molecular electric and magnetic
dipole moments in the sample. Then,
\begin{eqnarray}
{\bf E}_{\mbox{\tiny{ext}}}={\bf E}+\frac{N}{3\epsz}\bar{\bf p}_m,\,\,\,\,
{\bf B}_{\mbox{\tiny{ext}}}={\bf B}-\frac{2\muz N}{3}\bar{\bf m}_m
\end{eqnarray}

In the case of diamagnetic and paramagnetic molecules,
which are of considerable interest in NMR
experiments, the external fields experienced by a molecule
determine its microscopic (molecular) electric
and magnetic dipole moments. The mean moments of
all the molecules in the sample are linearly related to
the external field by the expressions
\begin{eqnarray}
\bar{\bf p}_m=\gamma_e\epsz{\bf E}_{\mbox{\tiny{ext}}},\,\,\,\,
\bar{\bf m}_m=\frac{\gamma_m}{\muz}{\bf B}_{\mbox{\tiny{ext}}}
\end{eqnarray}
where $\gamma_e$ is the average molecular polarizability and
$\gamma_m$ is the average molecular magnetizability; these
quantities are provided by the analysis of molecular
dynamics and have the dimensions of volume. (Note
that these $\gamma_e$, $\gamma_m$ parameters are not the magnetogyric
ratio that is usually denoted by this symbol in NMR
theory.) Substituting for the external fields from Eq.
[30] it is seen that the moments are aligned with the
macroscopic field:
\begin{eqnarray}
\bar{\bf p}_m=\frac{\epsz\gamma_e{\bf E}}{1-\gamma_e N/3},\,\,\,\,
\bar{\bf m}_m=\frac{(\gamma/\muz){\bf B}}{1+2\gamma_m N/3}
\end{eqnarray}
The polarization and magnetization are therefore
\begin{eqnarray}
{\bf P}&=& N\bar{\bf p}_m=\frac{\epsz\gamma_e N{\bf E}}{1-\gamma_e N/3},\nonumber\\
{\bf M}&=& N\bar{\bf m}_m=\frac{(\gamma N/\muz){\bf B}}{1+2\gamma_m N/3}\
\end{eqnarray}
and
\begin{eqnarray}
{\bf E}_{\mbox{\tiny{ext}}}=\frac{\bf E}{1-\gamma_e N/3},\,\,\,\,
{\bf B}_{\mbox{\tiny{ext}}}=\frac{\bf B}{1+2\gamma_m N/3},\,\,\,\,
\end{eqnarray}

\subsection*{Susceptibilities}
When calculating macroscopic fields, it is more usual
to introduce the susceptibilities $\chi_e$ and $\chi_m$, which are
defined by
\begin{equation*}
\chi_e=\frac{\gamma_e N}{1-\gamma_e N/3},\,\,\,\,
\chi_m=\frac{\gamma_m N}{1-\gamma_m N/3},\,\,\,\,
\end{equation*}
so that
\begin{eqnarray}
{\bf D}=\epsilon{\bf E},\,\,\,\,{\bf H}=\frac{1}{\mu}{\bf B}
\end{eqnarray}
where $\epsilon=\epsilon_0(1+\chi_e)$ and $\mu=\mu_0(1+\chi_m)$ are the
permittivity and permeability of the material, respectively.

Thus, in terms of the susceptibilities, P and M are
given by
\begin{eqnarray}
{\bf P}=\chi_e\epsz{\bf E},\,\,\,\,{\bf M}=\frac{\chi_m}{(1+\chi_m)}\frac{\bf B}{\muz}
\end{eqnarray}
hence,
\begin{eqnarray}
{\bf E}_{\mbox{\tiny{ext}}}=\left(1+\frac{\chi_e}{3}\right)\!{\bf E},\,\,\,
{\bf B}_{\mbox{\tiny{ext}}}=\left(1-\frac{2}{3}\frac{\chi_m}{(1+\chi_m)}\right)\!{\bf B}\,\,\,
\end{eqnarray}

The external fields can now be evaluated directly
from the macroscopic fields using Eq. [37], where the
macroscopic fields are found by solving the Maxwell
equations, Eq. [24], together with the constitutive
relations, Eq. [35].

\subsection*{Review}
So far, we have developed the theory of external fields
for both the electric and magnetic fields for several
reasons. First, it is based on a model of microscopic
charges for which the electric and magnetic fields are
strongly coupled because of the rapid motion of the
submolecular particles. Second, the macroscopic
fields remain coupled, but much more weakly because 
the macroscopic motions are much slower than the
microscopic ones and produce changes in the macroscopic
properties only over much longer time scales.
Third, the external fields can be analyzed systematically
when both the magnetic and electric fields are
varying slowly, the analysis leading to the remarkable
generalization of the sphere of Lorentz construct to
cases in which time variations are present. The results
of the construction may therefore be used when an
electric field is imposed on a sample in an NMR
experiment or when the applied magnetic field fluctuates.

We can therefore safely adopt the simplification
made in most NMR applications that the system is in
a macroscopically stationary state because any real
deviations from this state will not affect the manner in
which the external fields can be estimated. The advantage
of the stationary assumption is that the equations
governing the electric and magnetic fields are
then decoupled, allowing one to be treated independently
of the other. Therefore, in the examples that
follow to illustrate the theory attention is restricted to
the static magnetic field. Then, the calculation of the
external magnetic field from Eq. [37] requires a
knowledge of the expression for the macroscopic field
${\bf B}$. This is found via the macroscopic Maxwell equations;
the time-independent forms are
\begin{eqnarray*}
\nabla\cdot{\bf B}&\!=&\! 0,\,\,\,\,\nabla\times{\bf H}=0,\nonumber\\
{\bf B}&\!=&\!\mu{\bf H}=\muz(1+\chi){\bf H}=\muz({\bf H}+{\bf M})
\end{eqnarray*}
Their formal solution is treated in standard texts such
as (7, 8) and an outline in the context of NMR is given
elsewhere (e.g., 16) and in the appendix.

\section*{HOST FIELD}

\subsection*{General}
The macroscopic fields, and the fields from the
charged particles, are calculated using the approximation
in Eqs. [10] and [11], which is appropriate for
points at a large distance from the system of charges
that constitute each molecule. When calculating the
fields experienced by the nucleus of a host molecule,
there are contributions---the host fields---from the
system of charges constituting that molecule. This
situation requires us to evaluate the fields at an internal
point of the host molecule. Due to the proximity of
the electrical and magnetic field sources, these fields
will in general be more intense than those generated
by the other, distant, molecules. Their effect on the 
NMR resonance frequency of a nucleus, called the
chemical shift, is likely to be greater than that of the
external field, which produces the bulk susceptibility
shift. The host fields must therefore be determined
with care for each atomic nucleus in each molecular
species and it requires the full panoply of quantum
mechanics. This has been applied by many authors
from Ramsey (17) onward and is beyond the intended
scope of this article. However, the effect may usefully
be illustrated by a simple model.

\subsection*{Simple Model}
In the absence of an applied field, there will be no
preferred direction for a molecule and so the averaged
sources will have a distribution with spherical symmetry
about any fixed reference point, the nucleus in
the host molecule, for example. Suppose first that the
nucleus is at the center of a host atom so we assume
that the positive charge resides at the center. The
negative charge due to the averaged electron cloud
will appear as a spherical shell around it. In the
presence of an external field, the first-order perturbation
of the spherical shell will be a dipole term with its
axis in the direction of the applied field. Take the
preferred direction to be that of the unit vector ${\bf k}$. We
can take the nucleus as the origin of the coordinate
system without loss of generality. Then, the mean
charge density due to the electrons about the nucleus
can be written as
\begin{equation*}
\bar{\rho}_k({\bf x},t)=\rho_0(r,t)-\rho_1(r,t){\bf k}\cdot\frac{\bf x}{r}
\end{equation*}
where $r=|{\bf x}|$ is the distance of the point ${\bf x}$ from the
center. The charge density is here a negative quantity
and the center of the distribution is shifted in the
negative ${\bf k}$ direction due to the Lorentz force due to
the local applied field. The mean current density can
likewise be written
\begin{equation*}
\bar{\bf j}_k({\bf x},t)=j_1(r,t)\left({\bf k}\times\frac{\bf x}{r}\right)
\end{equation*}
Note that the spherically symmetrical current term
vanishes and the first-order term is a circular current
system about the axis ${\bf k}$. When $j_1$ is positive this
represents a paramagnetic effect; when negative, it
represents a diamagnetic effect.

The quasisteady microscopic host fields at the origin
are found by substituting these forms into the
appropriate terms of Eqs. [6] and [7] and performing
the integrations, giving
\begin{eqnarray*}
\bar{\bf e}_k(t)&=&\left(\frac{1}{3\epsz}\int\rho_1(r,t)dr\right)\!{\bf k},\nonumber\\
\bar{\bf b}_k(t)&=&\left(\frac{2\muz}{3}\int j_1(r,t)dr\right)\!{\bf k}
\end{eqnarray*}
We can also evaluate the average electric and magnetic
dipole moment of these distributions from
\begin{eqnarray*}
\bar{\bf p}_k(t)&=&-\left(\frac{4\pi}{3}\int\rho_1(r,t)r^3dr\right)\!{\bf k},\nonumber\\
\bar{\bf m}_k(t)&=&\left(\frac{4\pi}{3}\int j_1(r,t)r^3dr\right)\!{\bf k}
\end{eqnarray*}
The electric dipole is always oriented antiparallel to
the applied field and the magnetic dipole is parallel to
the applied field for paramagnetic molecules and antiparallel
for diamagnetic molecules. Hence,
\begin{eqnarray*}
\bar{\bf e}_k(t)&=&-\left(\frac{1}{4\pi\epsz}\frac{\int\rho_1 dr}{\int r^3\rho_1 dr}\right)
\!\bar{\bf p}_k(t),\nonumber\\
\bar{\bf b}_k(t)&=&-\left(\frac{\muz}{2\pi}\frac{\int j_1 dr}{\int r^3 j_1 dr}\right)\!\bar{\bf m}_k(t)
\end{eqnarray*}
The internal electric field is always antiparallel to the
electric dipole moment and the internal magnetic field
is always parallel to the magnetic dipole moment. The
exact relationship between the dipole moments and
the internal field, of course, depends on the electronic
structure of the molecule. The result can be written in
terms of the effective volumes of the molecule $V_e$ and
$V_m$, defined by
\begin{eqnarray}
V_e&\!=&\!4\pi\frac{\int r^3\rho_1 dr}{\int\rho_1 dr},\,\,\,\,\,\,\,\,\,\,
V_m=2\pi\frac{\int r^3 j_1 dr}{\int j_1 dr}\\
\bar{\bf e}_k(t)&\!=&\!-\frac{1}{\epsz V_e}\bar{\bf p}_k(t),\,\,\,\,
\bar{\bf b}_k(t)=-\frac{\muz}{V_m}\bar{\bf m}_k(t)
\end{eqnarray}
Similar expressions will therefore arise in the more
complicated cases that occur when the host atom is
chemically bound to a molecule. Averaging all possible
orientations of the molecule will produce charge
and current distributions with spherical symmetry
about the reference point, which is the nucleus of the
host atom. The external field will produce first-order
perturbations of these distributions that give rise to net
dipole moments and to the related host field at the
nucleus (center) of the host molecule.

\section*{DISCUSSION}

\subsection*{Sphere of Lorentz Argument}
We have shown above that the sphere of Lorentz
construct does indeed provide a means of estimating
the fields, both electrical and magnetic, at points
within a molecule embedded in a macroscopic sample
composed of other molecules. Our derivation does not
depend on an ad hoc ``hard" spherical construction but
it can be viewed as sort of ``soft" spherical one. The
analysis demands that we do not attempt to estimate
the fields experienced by an individual molecule, but
we estimate the average fields experienced by a large
collection of similar molecules so that both the dynamic
properties of the molecule and the locations of
the neighboring molecules are randomized. Under
these circumstances, the microscopic electromagnetic
sources in the close vicinity of the molecule display
spherical symmetry, as do the macroscopic ones. It is
the assumption of spherical symmetry that produces
the general results described by Eq. [28], not sphericity.
This explains why the radius of the assumed
sphere of Lorentz plays no role in the result; the
properties within a sphere of any radius drawn around
the center of a molecule will be symmetrical and lead
to the same result. Our argument is, however, not
independent of the distance scale so the size of the
averaging sphere is not totally irrelevant. The macroscopic
properties of the material are required to be
locally uniform, so the size of the macroscopic
smoothing volume must be chosen to ensure this
apparent homogeneity. This volume must be sufficiently
large to randomize the contribution of the least
abundant molecular species. Therefore, the smoothing
volume will be least when all the molecules in the
sample are of the same species. For a macroscopic
model to make sense, this volume must be less than
that of any macroscopic heterogeneities. Such cases
will be illustrated in the examples below.

\subsection*{Chemical Shift}
The magnetic field experienced by a nucleus in an
atom in a molecule is dependent upon the local
bonded structure of the atom. This field is the basis of
the experimentally important chemical shift of the
intrinsic resonance frequency away from the value
that pertains to the isolated atom. In what follows is
an explanation of this phenomenon and a model of a
molecular system that provides an estimate of the
order of magnitude of the chemical shift effect relative
to that of the BMS shift.

It is immaterial whether the host atom, in which the
nucleus of interest is located, is chemically bonded
into a larger molecule or not. In either case, the
nucleus will experience, on average, fields that are the
sum of the external fields, Eq. [28], and the host fields,
Eq. [39]. The size of the host fields will, of course,
depend on the molecular environment of the host
atom. Nevertheless, the dipole moments of the atom
or molecule to which it is bonded are linearly related
to the external fields as in Eq. [31] but with the mean
molecular polarizability and magnetizability replaced
by the quantities specific to the type of molecule in
question, $k$ say. Carrying out these substitutions gives
the host fields at the nucleus, Eq. [39]:
\begin{equation*}
\bar{\bf e}_k(t)=-\frac{\gamma_{e,k}}{V_e}{\bf E}_{\mbox{\tiny{ext}}},\,\,\,\,
\bar{\bf b}_k(t)=\frac{\gamma_{m,k}}{V_m}{\bf B}_{\mbox{\tiny{ext}}}
\end{equation*}
Then, we find the total fields at the nucleus to be
\begin{eqnarray*}
{\bf E}_k&\!=&\!{\bf E}_{\mbox{\tiny{ext}}}+\bar{\bf e}_k=\bigg{(}1+\frac{\chi_e}{3}\bigg{)}\!
\left(1-\frac{\gamma_{e,k}}{V_e}\right)\!{\bf E}\nonumber\\
{\bf B}_k&\!=&\!{\bf B}_{\mbox{\tiny{ext}}}+\bar{\bf b}_k=\left(1-\frac{2\chi_m}{3(1+\chi_m)}\right)\!
\left(1+\frac{\gamma_{m,k}}{V_m}\right)\!{\bf B}\nonumber\\
\end{eqnarray*}
where ${\bf E}$ and ${\bf B}$ are the macroscopic fields. These can
be put in a more revealing form by writing $\gamma_{e,k}$ in
terms of the mean polarizability $\gamma_e$ as
\begin{eqnarray}
\gamma_{e,k}=\frac{\gamma_{e,k}}{\gamma_e}\frac{\chi_e}{N(1+\chi_e/3)}
\end{eqnarray}
and similarly for the magnetizability. Then,
\begin{eqnarray}
{\bf E}_k&\!\!\!=&\!\!\!\bigg{(}1+\frac{\chi_e}{3}\bigg{)}\!\!\left(1-\frac{\gamma_{e,k}}{\gamma_e}
\frac{1}{N V_e}\frac{\chi_e}{(1+\chi_e/3)}\right)\!{\bf E}\\
{\bf B}_k&\!\!\!=&\!\!\!\bigg{(}1-\frac{2\chi_m}{3(1+\chi_m)}\bigg{)}\nonumber\\
&&\times\!\left(1+\frac{\gamma_{m,k}}{\gamma_m}
\frac{1}{N V_m}\frac{\chi_m}{(1+\chi_m/3)}\right)\!{\bf B}
\end{eqnarray}
Now, $N$ is the number of molecules per unit volume
so $1/NV$ is the ratio of the average intermolecular
volume to the volume of the molecule, as defined by
Eq. [38]. In fluids this ratio is, say, of the order of 10,
falling to unity when the molecules are densely
packed. The other factor is the ratio of the specific to
the mean value of $\gamma_e$ or $\gamma_m$. If there is only a single
species present, this ratio is unity so the chemical shift
due to the host field contribution will be an order of
magnitude greater than the bulk susceptibility shift
due to the external contribution. If the molecules are
diamagnetic then $\gamma_m<0$ in the host field and this
reduces the field experienced by the nucleus; so, the
effect is referred to as diamagnetic shielding. If the
molecule is paramagnetic, $\gamma_m>0$ and the nuclear
field is enhanced. When there are several species of
molecules present, the host contribution depends on
the properties of the molecule in question relative to
the average. It will be greatest when the molecule to
which the host atom is bonded is strongly paramagnetic
while the mean is weakly diamagnetic. These
considerations simply emphasize the need to have
accurate estimates of the host field at the site of the
nucleus of the host atom when the molecule is averaged
over all orientations. In this situation the required
analysis of chemical shifts is sophisticated and
computer intensive (e.g., 9).

\section*{ILLUSTRATIVE EXAMPLE}
In the following we illustrate the theory by evaluating
the external field and the susceptibility shift in a
suspension of RBCs following the procedure outlined
by Wolber et al. (18).

\subsection*{Susceptibility-Induced Shifts}
These authors begin their analysis with a system in
which a uniform strong magnetic field B0 is created in
a material of susceptibility $\chi_0$. There is then uniform
magnetization in the material given by Eq. [36]. If a
sample with susceptibility $\chi_s$ is now introduced then
the new field ${\bf B}$ will be the sum of the original field
and the field ${\bf B}$ due to the change in magnetization
\begin{eqnarray}
{\bf M}'=\frac{\chi_s{\bf B}}{\muz(1+\chi_s)}-\frac{\chi_0{\bf B}_0}{\muz(1+\chi_0)}
\end{eqnarray}
inside the sample and
\begin{eqnarray}
{\bf M}'=\frac{\chi_0({\bf B}-{\bf B}_0)}{\muz(1+\chi_0)}
\end{eqnarray}
outside the sample. But, ${\bf B}'$ also satisfies the following
equation (see appendix, Eq. [75]) for the case of a
single surface $S$:
\begin{eqnarray}
{\bf B}'=\muz{\bf M}'-\frac{\muz}{4\pi}\iint_S\frac{(\Delta{\bf M}'({\bf x}')\cdot{\bf n})({\bf x}-{\bf x}')}
{|{\bf x}-{\bf x}'|^3}d^2x'\,\,
\end{eqnarray}
where $S$ is the surface of the sample. As noted in the
appendix, these equations do not provide a full general
description of the field. For a full solution, it is
usual to solve the Laplace equation, Eq. [77], with
appropriate boundary conditions, for the scalar potential
(e.g., 16). However, Eq. [45] does provide a
useful {\emph approximate} form of the expression when the
magnetic susceptibilities are small. In practice, $\chi$ is of
the order of $-10\times 10^{-7}$ cgs-emu 
($-40\pi\times 10^{-10}$ SI units) so that 
$|{\bf B}'|/|{\bf B}_0|$ is also of order $-10\times 10^{-7}$.
Hence, to first order in small quantities we can write
\begin{eqnarray}
{\bf M}'\approx\frac{(\chi_s-\chi_0)}{\muz}{\bf B}_0
\end{eqnarray}
inside the sample and ${\bf M}'=0$ outside. Again, to first
order,
\begin{eqnarray}
{\bf B}'\approx(\chi_s&\!-&\!\chi_0){\bf B}_0\nonumber\\
&\!+&\!\frac{(\chi_s-\chi_0)}{4\pi}\iint_S\frac{({\bf B}_0\cdot{\bf n})({\bf x}-{\bf x}')}
{|{\bf x}-{\bf x}'|^3}d^2x'\,\,\,\,\,\,\,\,
\end{eqnarray}
The new macroscopic field inside the sample is thus,
approximately,
\begin{eqnarray}
{\bf B}({\bf x})=&\!{\bf B}_0&\!\!\!+(\chi_s-\chi_0)\nonumber\\
&\!\times&\!\!\!\!\left({\bf B}_0+\frac{1}{4\pi}\iint_S\frac{({\bf B}_0\cdot{\bf n})({\bf x}-{\bf x}')}
{|{\bf x}-{\bf x}'|^3}d^2x'\right)\,\,\,\,\,\,\,\,
\end{eqnarray}
The integral over the surface of the sample is a function
solely of the geometry of the surface. In general,
it is a function of the position x of the point within the
sample. For ellipsoidal surfaces (including the special
case of a spherical surface), however, the result is
independent of ${\bf x}$; in other words, the field within the
sample is uniform. For a sphere, 
\begin{eqnarray}
\frac{1}{4\pi}\iint_S\frac{({\bf B}_0\cdot{\bf n})({\bf x}-{\bf x}')}
{|{\bf x}-{\bf x}'|^3}d^2x'=-\frac{{\bf B}_0}{3}
\end{eqnarray}
so the macroscopic field is
\begin{eqnarray}
{\bf B}\approx\left(1+\frac{2}{3}(\chi_s-\chi_0\right)\!{\bf B}_0
\end{eqnarray}
In general, we may write
\begin{eqnarray}
\frac{1}{4\pi}\iint_S\frac{({\bf B}_0\cdot{\bf n})({\bf x}-{\bf x}')}
{|{\bf x}-{\bf x}'|^3}d^2x'=(\mathscr{D}_s-1){\bf B}_0
\end{eqnarray}
where $\mathscr{D}_s$ is the geometric demagnetizing factor that
is described by (18) among others; and, for a sphere
$\mathscr{D}_s=2/3$. For an infinite cylinder aligned at right
angles to the field, $\mathscr{D}_s=1/ 2$. Of course, there is no
surface effect for a cylinder aligned parallel to the
field and $\mathscr{D}_s=1$. A table of such factors is provided by
Chu et al. (5). The geometric factor is the same at all
points within these objects and in all objects with ellipsoidal
surfaces (16); the field is therefore uniform within
them. For other objects, the geometric factor will be a
function of position and the field will not be uniform. In
either case, we can write the macroscopic field as
\begin{eqnarray}
{\bf B}=(1+(\chi_s-\chi_0)\mathscr{D}_s){\bf B}_0
\end{eqnarray}
and the external field as
\begin{eqnarray}
{\bf B}_{\mbox{\tiny{ext}}}&\!=&\!\left(1-\frac{2\chi_s}{3(1+\chi_s)}\right)\!{\bf B}\nonumber\\
&\!\approx&\!{\bf B}_0+(\chi_s-\chi_0)\mathscr{D}_s{\bf B}_0-\frac{2}{3}\chi_s{\bf B}_0
\end{eqnarray}
to first order in the susceptibilities. This expression
describes the external field in absolute terms, i.e., it is
the field experienced at a point surrounded by a local
vacuum in the sample. The expression agrees with (5)
but it agrees with Eq. [1] in (18) only if the original
material is taken to be a vacuum, i.e., $\chi_0=0$.

The difference arises from the focus of the latter
authors on the frequency shift observed in NMR experiments.
This shift is governed by the {\emph change} in the
external field as a result of introducing the sample.
Before the introduction the macroscopic field is ${\bf B}_0$ so
the external field at the site of a nucleus is approximately
\begin{equation*}
{\bf B}_{\mbox{\tiny{ext}}}^1=\left(1-\frac{2}{3}\chi_0\right)\!{\bf B}_0
\end{equation*}
while the external field ${\bf B}_{\mbox{\tiny{ext}}}^2$
at the site after the introduction
of the sample is given by Eq. [53]. The
change is therefore
\begin{equation*}
\Delta{\bf B}_{\mbox{\tiny{ext}}}=\left(\mathscr{D}_s-\frac{2}{3}\right)\!(\chi_s-\chi_0){\bf B}_0
\end{equation*}
which is Eq. [1] in (18).

\subsection*{Sample Heterogeneity}
Wolber et al. (18) also consider the case in which the
sample is heterogeneous, consisting of plasma and
RBC. If the magnetic field in the sample is averaged
over a volume large enough to contain many erythrocytes,
the system can be considered to be uniform
with a volume-average susceptibility given by
\begin{eqnarray}
\chi_b=\frac{V_e\chi_e+V_p\chi_p}{V_e+V_p}
\end{eqnarray}
where the subscripts $e$ and $p$ denote erythrocyte and
plasma, respectively, and $V_e$ is the average volume of
an erythrocyte and $V_p$ the average volume of plasma
surrounding each erythrocyte. The macroscopic field
that determines the external field at a nucleus can then
be calculated as outlined above from the averages
bulk properties. If sample size is not large compared
to the size of the heterogeneities in the system, account
has to be taken of the detailed distribution of
erythrocytes in the neighborhood of the nucleus in
question. To do this, these authors conceptually surround
the site by a sphere, $\Sigma$ , that is comparable in
size to the averaging volume (see Fig. 7). 
\begin{figure}
\includegraphics[width=\columnwidth]{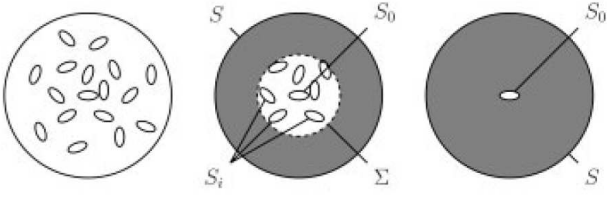}
\caption{Representation of a heterogeneous system of
RBCs as randomly dispersed spheroids; in reality, mammalian
RBCs are biconcave discs. The sample is bounded by
the surface $S$. In the second panel, the system is treated as
continuous outside the virtual sphere $\Sigma$ of Wolber et al.
(18). Within  the individual erythrocytes have surfaces $S_i$.
In the third panel, equivalent to the second, the system is
treated as continuous outside the selected RBC with surface $S_0$.}
\end{figure}
The field source outside this sphere contributes at the center an
approximate field, called the ``far" field, that is described by 
\begin{eqnarray}
{\bf B}'&\!\approx &\!
\frac{(\chi_b-\chi_0)}{4\pi}\iint_S\frac{({\bf B}_0\cdot{\bf n})({\bf x}-{\bf x}')}
{|{\bf x}-{\bf x}'|^3}d^2x'\nonumber\\
&\!&\!+\frac{(\chi_b-\chi_0)}{4\pi}\iint_{\Sigma}\frac{({\bf B}_0\cdot{\bf n})({\bf x}-{\bf x}')}
{|{\bf x}-{\bf x}'|^3}d^2x'\,\,
\end{eqnarray}
the normal being directed out of the sample in the first
integral and into the sphere in the second. As a result,
the two terms will almost cancel one another if the
averaging volume is comparable to the sample volume;
however, the purpose of the construction is to
place as much as possible of the sample outside the
surface $\Sigma$ so that it can be treated in an average
manner. Using the results from above, these expressions
yield
\begin{eqnarray}
{\bf B}' &\!\approx &\!(\chi_b-\chi_s)(\mathscr{D}_s-1){\bf B}_0+(\chi_s-\chi_0)\frac{1}{3}{\bf B}_0\nonumber\\
&\!=&\!(\chi_b-\chi_0)\left(\mathscr{D}_s-\frac{2}{3}\right)\!{\bf B}_0
\end{eqnarray}
This is the generalization of Eq. [3] of (18) for $\chi_0\ne 0$. 
To this must be added the ``near" field contributed
by sources within the sphere; this field is estimated by
taking an alternative macroscopic average over a volume
smaller than that of an erythrocyte but larger than
that of any internal structures. This construction ensures
that the susceptibilities $\chi_b$ and $\chi_p$ are uniform
within the erythrocyte and the surrounding plasma,
respectively. The macroscopic system is thus divided
into a hierarchy of {\emph macroscopic} systems, each of
which contribute to the macroscopic field at the site of
a nucleus. Then, the changes of magnetization within
the spherical volume, which are given by
\begin{eqnarray}
{\bf M}_e''\approx\frac{(\chi_e-\chi_0){\bf B}_0}{\muz},\,\,\,\,
{\bf M}_p''\approx\frac{(\chi_p-\chi_0){\bf B}_0}{\muz}
\end{eqnarray}
produce an approximate near macroscopic field at a
point in the plasma of
\begin{eqnarray}
{\bf B}_p''\approx &\!\!(&\! \!\!\!\!\chi_p-\chi_0){\bf B}_0\nonumber\\
&\!+&\!\!\frac{(\chi_p-\chi_e)}{4\pi}\sum_i\iint_{S_i}\frac{({\bf B}_0\cdot{\bf n}_i)({\bf x}-{\bf x}')}
{|{\bf x}-{\bf x}'|^3}d^2x'\,\,\,\,\,\,\,\,\,\\
&\!+&\!\!\frac{1}{4\pi}\iint_{\Sigma}\frac{(\chi_{\Sigma}-\chi_0)({\bf B}_0\cdot{\bf n})({\bf x}-{\bf x}')}
{|{\bf x}-{\bf x}'|^3}d^2x'
\end{eqnarray}
Here, the sum is taken of the integrals over the surfaces
$S_i$ of all the erythrocytes within the sphere with
the normal at the surface ${\bf n}_i$ pointing into the erythrocyte;
and, the final term is the integral over the
surface of the surrounding sphere $\Sigma$ with outwardly
directed normal. The susceptibility $\chi_{\Sigma}$ at points on
this sphere will vary according to whether the point
lies in plasma or erythrocyte. However, the sphere
will, by construction, intersect a large number of
erythrocytes and so the integral may be approximated
by replacing  by the average bulk susceptibility $\chi_b$
on the surface. In this approximation
\begin{eqnarray}
\frac{1}{4\pi}\iint_{\Sigma}\frac{(\chi_{\Sigma}-\chi_0)({\bf B}_0\cdot{\bf n})({\bf x}-{\bf x}')}
{|{\bf x}-{\bf x}'|^3}d^2x'\nonumber\\
\approx -(\chi_b-\chi_0)\frac{1}{3}{\bf B}_0
\end{eqnarray}
which, of course, cancels the contribution of the surface
of the sphere to the field ${\bf B}'$.

The near contribution to the macroscopic field in
an RBC is, in the same approximation, given by
\begin{eqnarray}
{\bf B}_e''\approx &\!\!(&\! \!\!\!\!\chi_e-\chi_0){\bf B}_0\nonumber\\
&\!+&\!\!\frac{(\chi_e-\chi_p)}{4\pi}\iint_{S_0}\frac{({\bf B}_0\cdot{\bf n}_0)({\bf x}-{\bf x}')}
{|{\bf x}-{\bf x}'|^3}d^2x'\\
&\!+&\!\!\frac{(\chi_e-\chi_p)}{4\pi}\sum_{i=1}^n\iint_{S_i}\frac{({\bf B}_0\cdot{\bf n}_i)({\bf x}-{\bf x}')}
{|{\bf x}-{\bf x}'|^3}d^2x'\,\,\,\,\,\,\,\,\,\,\,\\
&\!+&\!\!\frac{1}{4\pi}\iint_{\Sigma}\frac{(\chi_{\Sigma}-\chi_0)({\bf B}_0\cdot{\bf n})({\bf x}-{\bf x}')}
{|{\bf x}-{\bf x}'|^3}d^2x'
\end{eqnarray}
where $S_0$ is the surface of the selected RBC and the
sum is taken over all the others ($n$ in total) in the
sphere, the normal now pointing out of each RBC.
The first integral is evaluated at an internal point and
can be treated as before in Eq. [51], replacing s by
$\mathscr{D}_e$, which is the geometric factor appropriate to the
shape of the RBC. The integral over the sphere $\Sigma$ is
treated as before but it appears to have been omitted
by (18). After making this correction, the total macroscopic
field at a point in an RBC is
\begin{eqnarray}
{\bf B}\approx &\!\!\!\big{(}&\!\!\!\!\!1+(\chi_e-\chi_0)+(\chi_b-\chi_0)(\mathscr{D}_s-1)\nonumber\\
&\!+&\!\!(\chi_e-\chi_p)(\mathscr{D}_e-1)\big{)}{\bf B}_0+(\chi_e-\chi_p){\bf B}'''\,\,
\end{eqnarray}
where
\begin{eqnarray}
{\bf B}'''=
\frac{1}{4\pi}\sum_{i=1}^n\iint_{S_i}\frac{({\bf B}_0\cdot{\bf n}_i)({\bf x}-{\bf x}')}
{|{\bf x}-{\bf x}'|^3}d^2x'
\end{eqnarray}
To first order in the susceptibilities, the external field
experienced by a nucleus in an RBC will be
\begin{eqnarray}
{\bf B}_{\mbox{\tiny{ext}}}
&\!\!\! \approx &\!\!\!\bigg{(}1+(\chi_e-\chi_0)+(\chi_b-\chi_0)(\mathscr{D}_s-1)\nonumber\\
&\!\!\!\!\!\!\!\!\!\!\!\!+&\!\!\!\!\!\!\!\!(\chi_e-\chi_p)(\mathscr{D}_e-1)-\frac{2}{3}\chi_e\bigg{)}{\bf B}_0+(\chi_e-\chi_p){\bf B}'''\,\,\,\,\,\,\,\,\,\,
\end{eqnarray}
This expression corrects and generalizes Eq. [5] in
(18). These authors introduced a further term to describe
the contribution of the particles within a sphere
of Lorentz drawn about a nucleus within the RBC;
then, they argued that it vanished ``because of symmetry."
We have shown above that the contribution is
in fact included in the estimate of the external field if
the particles are randomized within the ``local" macroscopic
volume that surrounds the nucleus, so the
final result is the same.

Wolber et al. (18) then claim that ${\bf B}'''$ vanishes due
to symmetry. However, in general, there will be no
symmetry in the distribution of the other RBCs about
the selected cell. An exact evaluation is then impossible.
We argue that this contribution can be estimated
only if the experiment samples enough RBCs (either
in volume or time) that the configuration external to
any individual RBC is randomized. Then, the structured
medium can be replaced by a homogeneous
medium with a mean susceptibility given by Eq. [54].
In this case
\begin{eqnarray*}
{\bf B}_e''\approx &\!\!\!(&\! \!\!\!\!\chi_e-\chi_0){\bf B}_0\nonumber\\
&\!+&\!\!\frac{(\chi_e-\chi_b)}{4\pi}\iint_{S_0}\frac{({\bf B}_0\cdot{\bf n}_0)({\bf x}-{\bf x}')}
{|{\bf x}-{\bf x}'|^3}d^2x'\,\,\,\,\,\,\,\,\,\\
&\!+&\!\!\frac{(\chi_b-\chi_0)}{4\pi}\iint_{\Sigma}\frac{({\bf B}_0\cdot{\bf n})({\bf x}-{\bf x}')}
{|{\bf x}-{\bf x}'|^3}d^2x'\\
&&\!\!\!\!\!\!\!\!\!\!\!\!\!\!\!\!\!=\,
(\chi_e-\chi_0){\bf B}_0+\frac{(\chi_e-\chi_b)}{4\pi}(\mathscr{D}_e-1){\bf B}_0\nonumber\\
&\!-&\!\!(\chi_b-\chi_0)\frac{1}{3}{\bf B}_0
\end{eqnarray*}
and the external field becomes
\begin{eqnarray}
{\bf B}_{\mbox{\tiny{ext}}}
&\!\!\! \approx &\!\!\!\bigg{(}1+(\chi_e-\chi_0)+(\chi_b-\chi_0)(\mathscr{D}_s-1)\nonumber\\
&+&\!\!(\chi_e-\chi_b)(\mathscr{D}_e-1)-\frac{2}{3}\chi_e\bigg{)}{\bf B}_0
\end{eqnarray}
Of course, this is exactly the same result as that which
would be obtained by ignoring the first spherical
construction and applying Eq. [75] directly to the
surfaces that define the selected erythrocyte and the
sample volume. The selected RBC is, on average,
completely surrounded not by plasma alone but by
plasma containing other RBCs, so the susceptibility in
the surrounding medium is $\chi_b$, not $\chi_p$. If we set $\chi_0=0$, 
we obtain
\begin{eqnarray}
{\bf B}_{\mbox{\tiny{ext}}}
&\!\!\! \approx &\!\!\!\bigg{(}1+\chi_b(\mathscr{D}_s-1)\nonumber\\
&+&\!\!(\chi_e-\chi_b)(\mathscr{D}_e-1)+\frac{1}{3}\chi_e\bigg{)}{\bf B}_0
\end{eqnarray}
which corrects the result given by (16),
\begin{eqnarray}
{\bf B}_{\mbox{\tiny{ext}}}&\!\!=&\!\nonumber\\
\Big{(}&\!\!\!1&\!\!\!+\chi_b\left(\mathscr{D}_s-\frac{2}{3}\right)
+(\chi_e-\chi_p)\!\left(\mathscr{D}_e-\frac{2}{3}\right)\!\Big{)}\!{\bf B}_0\,\,\,\,\,\,\,\,\,
\end{eqnarray}

\section*{CONCLUSIONS}

In conclusion, we can finally consider the validity of
the assumption of spherical symmetry for a molecule
averaged over all its orientations: Exact spherical
symmetry would result if there were no preferred
directions. However, applied electric or magnetic
fields do provide preferred directions and the molecules
(especially macromolecules) may be distorted
{\emph systematically} as a result. These departures from
spherical symmetry will lead to a second-order correction
being required to the estimate of the external
field. But, the first-order effect estimated above is of
order  $\chi\approx 10^{-7}$, so second-order effects will be of
order $\chi^2\approx 10^{-14}$ so they will be negligible. The
expressions in Eq. [28] therefore provide accurate
estimates of the averaged external fields. Hence, we
have shown that the mathematical constructs presented
above are well defined and allow the bulk
susceptibility shift to be calculated on a more rigorous
basis than hitherto.

In practice, an NMR experiment samples a large
number of nuclei in a macroscopic volume over a
macroscopic time interval, and each will experience a
fluctuating environment as the neighboring molecules
move around. As a result, the molecules will not
experience a single average field but a spread of field
strengths. This fluctuation will appear as a broadening
of the resonance line about the mean value that can be
calculated as described here. In principle, the size of
the fluctuations can be estimated from the width at
half height of the NMR spectral line if the geometric
factors are known accurately (see previous accompanying
article).

\section*{APPENDIX}

The steady-state solution of the Maxwell equations
that describes the magnetic field in a sample is given
by the first term in Eq. [22]. Recalling the fact that the
integration is performed over all space, this expression
can be transformed into two alternative forms by
integrating by parts and assuming that the surface
integral vanishes sufficiently far from the sample nuclei.
These expressions are
\begin{eqnarray}
{\bf B}&\!\!=&\!\!\nabla\times\iiint\frac{\muz}{4\pi}\frac{{\bf M}({\bf x}')\times({\bf x}-{\bf x}')}
{|{\bf x}-{\bf x}'|^3}d^3x'\nonumber\\
&\!\!=&\!\!\muz{\bf M}({\bf x})-\nabla\!\left(\frac{\muz}{4\pi}\iint\frac{{\bf M}({\bf x}')\cdot({\bf x}-{\bf x}')}
{|{\bf x}-{\bf x}'|^3}d^3x'\right)\,\,\,\,\,\,\,\,
\end{eqnarray}
These two forms are equivalent for arbitrary volumes.

The first form displays the construction of the field
from the vector potential
\begin{eqnarray*}
{\bf A}=\iiint\frac{\muz}{4\pi}\frac{{\bf M}({\bf x}')\times({\bf x}-{\bf x}')}
{|{\bf x}-{\bf x}'|^3}d^3x'
\end{eqnarray*}
because the Maxwell equation $\nabla\cdot{\bf B}=0$ guarantees
that we can write ${\bf B}=\nabla\times{\bf A}$.

The second can be rewritten as
\begin{eqnarray}
\frac{1}{\muz}{\bf B}-{\bf M} &\!\!=&\!\!{\bf H}\nonumber\\
&\!\!=&\!\!\nabla\!\left(\frac{1}{4\pi}\iiint\frac{{\bf M}({\bf x}')\cdot({\bf x}-{\bf x}')}
{|{\bf x}-{\bf x}'|^3}d^3x'\right)\,\,\,\,\,\,
\end{eqnarray}
which exhibits the construction from the {\emph scalar} potential
\begin{eqnarray*}
\Phi_M=\iiint\frac{{\bf M}({\bf x}')\cdot({\bf x}-{\bf x}')}
{4\pi|{\bf x}-{\bf x}'|^3}d^3x'
\end{eqnarray*}
because the Maxwell equation $\nabla\times{\bf H}=0$ guarantees
that we can write ${\bf H}=-\nabla\Phi_M$.

If $M$ is everywhere differentiable and the integral
is taken over all space, the scalar potential can be
rewritten as
\begin{eqnarray}
\Phi_M=-\iiint\frac{\nabla_{{\bf x}'}\!\cdot{\bf M}}{4\pi|{\bf x}-{\bf x}'|}d^3x'
\end{eqnarray}
which is the solution over all space of the Poisson
equation 
\begin{eqnarray}
\nabla^2\Phi_m=\nabla\cdot{\bf M}
\end{eqnarray}

The Poisson equation is just a form of the wave
equations [4] and [5] when there is no time dependence.
The solutions can therefore be obtained from
the solutions of the wave equation, Eqs. [6] and [7],
by performing the trivial integration over time when
the integrand has no time dependence. This produces
Eq. [72].

If space contains media with discontinuous distributions
of ${\bf M}$, the field may be determined from either
of the two equivalent forms of Eq. [70], but the
volume integrals need to be evaluated with some care.

Suppose space is divided into $n$ regions with volumes
$V_i$ within which ${\bf M}_i(x)$ is differentiable. Let $V_i$
and $V_j$ have a common surface $S_{ij}$, which may have
zero extent (Fig. 8).
\begin{figure}
\includegraphics[width=70mm]{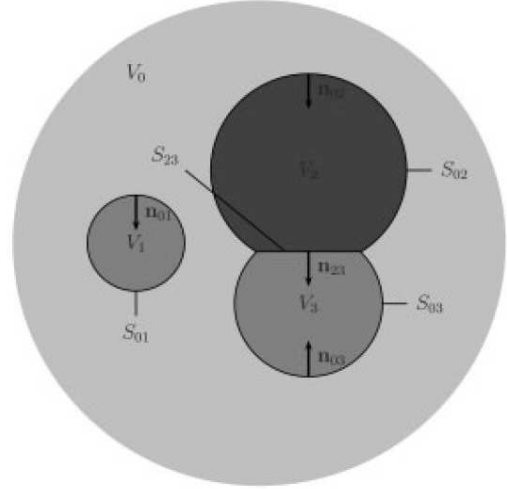}
\caption{Heterogeneous system with volumes $V_i$ embedded
in a volume $V_0$. The surfaces separating volumes $V_i$ and
$V_j$ are denoted $S_{ij}$ and the normal vector ${\bf n}_{ij}$ 
at points of the surface $S_{ij}$ is directed from 
$V_i$ to $V_j$, where $j>i$.}
\end{figure}
Then, we can transform the
vector potential to obtain a contribution from both
within each volume $V_i$ and from each surface $S_{ij}$
\begin{eqnarray*}
{\bf A}=\sum_{i=1}^n &&\!\!\!\!\!\! \frac{\muz}{4\pi}\Bigg{(}\sum_{j=1}^n\!'
\iint_{S_{ij}}\frac{{\bf M}_i({\bf x}'-)\times{\bf n}_{ij}}{|{\bf x}-{\bf x}'|}d^2x'\nonumber\\
&& +\iiint\frac{\nabla_{{\bf x}'}\!\times{\bf M}_i({\bf x}')}{|{\bf x}-{\bf x}'|}d^3x'\Bigg{)}
\end{eqnarray*}
Here, the magnetization ${\bf M}_i({\bf x}'-)$ in the surface integral
is evaluated just inside the volume $V_i$, the primed
sum excludes the term with $i=j$, and ${\bf n}_{ij}$ is the
outward normal from $V_i$ to $V_j$ on $S_{ij}$.

If the magnetization is uniform within $V_i$,  
$\nabla\times{\bf M}_i=0$ within $V_i$ and the volume 
integral vanishes. If
the medium is uniform so $\chi_i$ is constant within $V_i$,
then $\nabla\times{\bf M}_i=\chi_i\nabla\times{\bf H}$ 
within $V_i$. In the magnetostatic
case, the volume integral will again vanish.

If space is composed only of such regions
\begin{equation*}
{\bf A}=\frac{\muz}{4\pi}\sum_{i=1}^n\sum_{j=1}^n\!'\iint_{S_{ij}}\frac{{\bf M}_i({\bf x}'-)\times{\bf n}_{ij}}
{|{\bf x}-{\bf x}'|}d^2x'
\end{equation*}
and the terms can be paired because $S_{ij}=S_{ji}$ and
${\bf n}_{ij}=-{\bf n}_{ji}$ to give
\begin{eqnarray*}
\!\!\!\!{\bf A}&&\!\!\!\!\!=\nonumber\\
\frac{\muz}{4\pi}&&\!\!\!\!\!\sum_{i=1}^n\sum_{j=i+1}^n\!\!'\iint_{S_{ij}}\!
\frac{({\bf M}_i({\bf x}'-)-{\bf M}_j({\bf x}'+))\times{\bf n}_{ij}}
{|{\bf x}-{\bf x}'|}d^2x'\,\,\,\,\,\,\,\,
\end{eqnarray*}
where now ${\bf M}_j({\bf x}'+)$ is now evaluated just outside the
volume $V_i$. The quantity 
$\Delta{\bf M}_{ij}({\bf x}')={\bf M}_j({\bf x}'+)-{\bf M}_i({\bf x}'-)$
is the jump in magnetization as the boundary
$S_{ij}$ is crossed from $V_i$ to $V_j$ (in the direction of 
${\bf n}_{ij})$ at the point ${\bf x}'$.

The field follows immediately:
\begin{eqnarray}
{\bf B}=-\frac{\muz}{4\pi}&&\!\!\!\!\!\sum_{i=1}^n\sum_{j=i+1}^n\nonumber\\
\times&&\!\!\!\!\!\!\iint_{S_{ij}}\!
\frac{(\Delta{\bf M}_{ij}({\bf x}')\times{\bf n}_{ij})\times({\bf x}-{\bf x}')}
{|{\bf x}-{\bf x}'|^3}d^2x'\,\,\,\,\,\,\,\,\,\,
\end{eqnarray}

Alternatively, we may work from a similar version
of the scalar potential:
\begin{eqnarray*}
\Phi_M=\sum_{i=1}^n\frac{1}{4\pi}\Bigg{(}\sum_{j=1}^n\!'\iint_{S_{ij}}
\frac{{\bf M}_i({\bf x}'-)\cdot{\bf n}_{ij}}{|{\bf x}-{\bf x}'|}d^2x'\nonumber\\
-\iiint\frac{\nabla_{{\bf x}'}\!\cdot{\bf M}_i({\bf x}')}{|{\bf x}-{\bf x}'|}d^3x'\Bigg{)}
\end{eqnarray*}

Under the same conditions as before  $\nabla\cdot{\bf M}=0$ and
the volume integrals vanish, leaving
\begin{eqnarray*}
\Phi_M=-\frac{1}{4\pi}\sum_{i=1}^n\sum_{j=I+1}^n\!\!'\iint_{S_{ij}}
\frac{\Delta{\bf M}_{ij}({\bf x}')\cdot{\bf n}_{ij}}{|{\bf x}-{\bf x}'|}d^2x'
\end{eqnarray*}

The magnetic field is now given by
\begin{eqnarray}
{\bf B}&&\!\!\!\!\! = \muz{\bf M}\nonumber\\
-&&\!\!\!\!\!\frac{\muz}{4\pi}\sum_{i=1}^n\sum_{j=i+1}^n\!\iint_{S_{ij}}\!\!
\frac{(\Delta{\bf M}_{ij}({\bf x}')\cdot{\bf n}_{ij})({\bf x}-{\bf x}')}
{|{\bf x}-{\bf x}'|^3}d^2x'\,\,\,\,\,\,\,\,\,\,
\end{eqnarray}
where ${\bf M}$ is the appropriate magnetization at the point ${\bf x}$.

In this form it is easier to implement the boundary
conditions when the materials are diamagnetic or
paramagnetic. In either case, ${\bf M}\cdot{\bf n}$ is the normal
component of the magnetization at the boundary and
\begin{equation*}
\Delta{\bf M}_{ij}\cdot{\bf n}=\frac{(\chi_j-\chi_i)}{\muz(1+\chi_j)(1+\chi_i)}
{\bf B}({\bf x}')\cdot{\bf n}
\end{equation*}
because Maxwell's equations require the normal component
of ${\bf B}$ to be continuous across a boundary.

Hence,
\begin{eqnarray}
{\bf B}&&\!\!\!\!\!=\frac{\chi}{1+\chi}{\bf B}-\frac{1}{4\pi}\sum_{i=1}^n\sum_{j=i+1}^n\nonumber\\
\times&&\!\!\!\!\!\!\!\iint_{S_{ij}}\!\frac{(\chi_j-\chi_i)}{(1+\chi_j)(1+\chi_i)}
\frac{({\bf B}({\bf x}')\cdot{\bf n}_{ij})({\bf x}-{\bf x}')}{|{\bf x}-{\bf x}'|^3}d^2x'\,\,\,\,\,\,\,\,\,
\end{eqnarray}
This appears to be an equation that defines ${\bf B}$ implicitly
everywhere but it cannot be implemented as such
because it does not define the discontinuous tangential
component of ${\bf B}$ fields on the surfaces $S_{ij}$. The general
method of solution when $\nabla\cdot{\bf M}= 0$ everywhere is to
solve Eq. [73] with vanishing right side:
\begin{eqnarray}
\nabla^2\Phi_M=0
\end{eqnarray}
This is Laplace's equation, a version of Poisson's
equation without sources. Because there are no
sources, the construction in Eq. [72] is not applicable
and the solution of Laplace's equation must be determined
from the conditions imposed on the boundaries.
The boundary conditions on $\Phi_M$ at the surfaces
between the different media are fully determined by
the Maxwell equations, namely, that $\Phi_M$ is continuous
across the boundary so that the components of ${\bf H}$
parallel to the boundary are continuous and the component
of ${\bf B}$ normal to the boundary is continuous, i.e.,
$(1+\chi)({\bf n}\cdot\nabla\Phi_M)$ is continuous.

These prescriptions allow the magnetic field ${\bf B}$ to
be found exactly in all circumstances. In practice,
however, the smallness of the susceptibilities allows
approximations to be made that greatly simplify the
estimation of the field. This is demonstrated in the
illustrative example in the text above.

\section*{REFERENCES}
\begin{enumerate}
\item Brindle KM, Brown FF, Campbell ID, Grathwohl C,
Kuchel PW. Application of spin echo nuclear magnetic
resonance to whole cell systems: membrane transport.
Biochem J 1979; 180:37--44. 
\item Endre ZH, Chapman BE, Kuchel PW. Cell volume
dependence of 1H spin echo NMR signals in human
erythrocyte suspensions: the influence of in-situ field
gradients. Biochem Biophys Acta 1984; 803:137--144.
\item Kirk K, Kuchel PW. The contribution of magnetic susceptibility
effects to transmembrane chemical shift differences
in the $^{31}$P NMR spectra of oxygenated erythrocyte
suspensions. J Biol Chem 1989; 263:30--134.
\item Matwiyoff NA, Gasparovic C, Mazurchuk R,
Matwiyoff G. The line-shapes of the water proton resonances 
of red-blood-cells containing carbonyl hemoglobin,
deoxyhemoglobin, and methemoglobin---implications
for the interpretation of proton MRI at fields of
1.5-T and below. Magn Reson Imag 1990; 8:295--301.
\item Chu SC-K, Xu Y, Balschi JA, Springer CS. Bulk magnetic-
susceptibility shifts in NMR-studies of compartmentalized
samples: use of paramagnetic reagents.
Magn Reson Med 1990; 13:239--262.
\item Levitt MH. Demagnetization field effects in two-dimensional
solution NMR. Concepts Magn Reson 1996;
8:77--103.
\item Jackson JD. Classical Electrodynamics, 3rd Ed. New
York: John Wiley \& Sons; 1999.
\item Shadowitz A. The Electromagnetic Field. New York:
Dover; 1975 (reprinted).
\item Springer CS. Physicochemical principles influencing
magnetopharmaceuticals. In: Gillies RJ, ed. NMR in
Physiology and Biomedicine. San Diego: Academic
Press; 1994. p 75--99.
\item Lorentz HA. The Theory of Electrons and Its Applications
to the Phenomena of Light and Radiant Heat, 2nd
Ed. New York: Dover; 1915 (reprinted 1942).
\item Feynman RP, Leighton RB, Sands M. The Feynman
Lectures on Physics. London: Addison-Wesley; 1964.
\item Dickinson WC. The time average magnetic field at the
nucleus in nuclear magnetic resonance experiments.
Phys Rev 1951; 81:717--731.
\item Ramsey NF. Chemical effects in nuclear magnetic resonance
and diamagnetic susceptibility. Phys Rev 1952;
86:243--246.
\item Schey HM. Div, Grad and All That. New York: Norton;
1997.
\item Russakoff G. A derivation of the macroscopic Maxwell
equations. Am J Phys 1970; 38:1188--1195.
\item Kuchel PW, Bulliman BT. Perturbation of homogeneous
magnetic fields by isolated cells modelled as
single and confocal spheroids: implications for magnetic
resonance spectroscopy and imaging. NMR
Biomed 1989; 2:151--160.
\item Ramsey NF. Magnetic shielding of nuclei in molecules.
Phys Rev 1950; 78:699--703.
\item Wolber J, Cherubini A, Leach MO, Bifone A. Hyperpolarized
$^{129}$Xe NMR as a probe for blood oxygenation.
Magn Reson Med 2000; 43:491--496.
\end{enumerate}

\end{document}